\documentclass[format=acmsmall, review=false, screen=true]{acmart}

\usepackage{booktabs} 

\usepackage{amsmath} 
\usepackage{amsfonts} 
\usepackage{amssymb}

\usepackage{subfigure}
\usepackage{pgfplots}
\usepackage{amsmath,tabularx}
\usepgfplotslibrary{groupplots}
\usepackage[ruled]{algorithm2e} 

\SetAlFnt{\small}
\SetAlCapFnt{\small}
\SetAlCapNameFnt{\small}
\SetAlCapHSkip{0pt}
\IncMargin{-\parindent}

\usepackage{tabularx,booktabs}

\newcolumntype{C}{>{\centering\arraybackslash}X} 
\newcolumntype{L}{>{\raggedright\arraybackslash}X}

\makeatletter
\renewcommand\@makefntext[1]{%
	\parindent 1em%
	\noindent
	\hbox{\hss\@makefnmark}#1}
\makeatother

\makeatletter
\newcommand\footnoteref[1]{\protected@xdef\@thefnmark{\ref{#1}}\@footnotemark}
\makeatother

\usepackage{tikz}
\usetikzlibrary{patterns}

\newcommand{\as}[1]{\textcolor{blue}{$^{\textrm{}}${#1}}}

\usepackage{lipsum}
\usepackage{soul}
\usepackage{multirow}

\begin{document}
\title[Polarization and Fake News: Early Warning of Potential Misinformation Targets]{Polarization and Fake News: Early Warning of Potential Misinformation Targets}

\author{Michela Del Vicario}
\affiliation{%
  \institution{IMT School for Advanced Studies Lucca}
  \streetaddress{Piazza S. Ponziano, 6}
  \city{Lucca}
  \postcode{55100}
  \country{Italy}}

\author{Walter Quattrociocchi}
\affiliation{%
  \institution{Ca' Foscari University of Venice}
  \streetaddress{Via Torino, 155}
    \postcode{30172}
  \city{Venice}
  \country{Italy}
}
\author{Antonio Scala} 
\affiliation{%
 \institution{ISC-CNR Sapienza University of Rome}
   \streetaddress{Via dei Taurini, 19}
     \postcode{00185}
 \city{Rome} 
 \country{Italy}}

\author{Fabiana Zollo}
\affiliation{%
  \institution{Ca' Foscari University of Venice}
    \streetaddress{Via Torino, 155}
      \postcode{30172}
  \city{Venice}
  \country{Italy}
}

\begin{abstract}
Users polarization and confirmation bias play a key role in misinformation spreading on online social media. 
Our aim is to use this information to determine in advance potential targets for hoaxes and fake news. In this paper, we introduce a general framework for promptly identifying polarizing content on social media and, thus, ``predicting'' future fake news topics. We validate the performances of the proposed methodology on a massive Italian Facebook dataset, showing that we are able to identify topics that are susceptible to misinformation with 77\% accuracy. Moreover, such information may be embedded as a new feature in an additional classifier able to recognize fake news with 91\% accuracy. The novelty of our approach consists in taking into account a series of characteristics related to users behavior on online social media, making a first, important step towards the smoothing of polarization and the mitigation of misinformation phenomena. 
\end{abstract}

%
%


%
%

\keywords{social media, fake news, misinformation, polarization, classification}

\thanks{The authors acknowledge financial support from IMT/Extrapola Srl project. The funders had no role in study design, data collection and analysis, decision to publish, or preparation of the manuscript.}

\maketitle
 
\renewcommand{\shortauthors}{M. Del Vicario et al.}

\section{Introduction}

As of the third quarter of 2017, Facebook had 2.07 billion monthly active users~\cite{statista2018}, leading the rank of most popular social networking sites in the world. In the meantime, Oxford Dictionaries announced ``post-truth'' as the 2016 international Word of the Year~\cite{oxford}. Defined as an adjective ``relating to or denoting circumstances in which objective facts are less influential in shaping public opinion than appeals  to emotion and personal belief'', the term has been largely used in the context of the Brexit and Donald Trump's election in the United States and benefited from the rise of social media as news source. Indeed, Internet changed the process of knowledge production in an unexpected way. The advent of social media and microblogging platforms has revolutionized the way users access content, communicate and get informed. People can access to an unprecedented amount of information --only on Facebook more than 3M posts are generated per minute~\cite{allen2017what}-- without the intermediation of journalists or experts, thus actively participating in the diffusion as well as the production of content. Social media have rapidly become the main information source for many of their users: over half (51\%) of US users now get news via social media~\cite{newman2017reuters}. However, recent studies found that confirmation bias --i.e., the human tendency to acquire information adhering to one's system of beliefs-- plays a pivotal role in information cascades~\cite{del2016spreading}. Selective exposure has a crucial role in content diffusion and facilitates the formation of echo chambers --groups of like-minded people who acquire, reinforce and shape their preferred narrative~\cite{schmidt2017anatomy,delvicario2017mapping}. In this scenario, dissenting information usually gets ignored~\cite{zollo2017debunking}, thus the effectiveness of debunking, fact-checking and other similar solutions turns out to be strongly limited. 

Since 2013 the World Economic Forum (WEF) has been placing the global danger of massive digital misinformation at the core of other technological and geopolitical risks~\cite{howell2013digital}. Hence, a fundamental scientific challenge is how to support citizens in gathering trustworthy information to participate meaningfully in public debates and societal decision making. However, attention should be paid: since the problem is complex, solutions could prove to be wrong and disastrous. For instance, relying on machine learning algorithms alone (and scientists behind) to separate the truth from the false is na\"if and dangerous, and might have severe consequences.

As far as we know, misinformation spreading on social media is directly related to the increasing polarization and segregation of users~\cite{del2016spreading,zollo2017debunking,quattrociocchi2016echo,zollo2017misinformation}. Given the key role of confirmation bias in fostering polarization, our aim is to use the latter as a proxy to determine in advance the targets for hoaxes and fake news.
In this paper, we introduce a general framework for identifying polarizing content on social media in a timely manner --and, thus, ``predicting'' future fake news topics. We validate the performances of the proposed methodology on a massive Italian Facebook dataset with more than 300K news from official newspapers and  50K posts from websites disseminating either fake or unsubstantiated information. However, the framework is easily extensible to other social networks and microblogging platforms. Our results show that we are able to identify polarizing topics with 77\% accuracy (0.73 AUC). Our approach would be of great importance to tackle misinformation spreading online, and could represent a key element of a system (\textit{observatory}) to constantly monitor information flow in real time, allowing to issuing a warning about topics that require special caution. Moreover, we show that the output of our framework --i.e., whether a topic is susceptible to misinformation-- may also be used as a new feature in a classifier able to recognize fake news with 91\% accuracy (0.94 AUC).

Despite the goodness of our results, we are aware of the limits of this approach. Indeed, in spite of the great benefits w.r.t. pure misinformation, the identification of disinformation or propaganda has to be tackled with due caution. However, the novelty of our approach consists in taking into account a series of characteristics related to users behavior and polarization, making a first, important step towards the mitigation of misinformation spreading on online social media. 

The manuscript is structured as follows: in Section~\ref{relatedwork} we provide an overview of the related work; in Section~\ref{framework} we introduce our framework for the early warning, and focus on some interesting insights about users behavior w.r.t. controversial content; in Section~\ref{usecase} we describe a real use-case of the framework on Facebook data; in Section~\ref{fakedetection} we show how information provided by our framework may be exploited for fake news detection and classification; finally, we draw some conclusions in Section~\ref{conclusion}.

\section{Related Work} \label{relatedwork}
A review of previous literature reveals a series of works aiming at detecting misinformation on Twitter, ranging from the identification of suspicious or malicious behavioral patterns by exploiting supervised learning techniques~\cite{antoniadis2015model,rajdev2015fake}, to automated approaches for spotting and debunking misleading content~\cite{boididou2017learning,boididou2017verifying,popescu2010detecting}, to the assessment of the ``credibility'' of a given set of tweets~\cite{castillo2011information}. A complementary line of research focused on similar issues on other platforms, trying to identify hoax articles on Wikipedia~\cite{kumar2016disinformation}, study users' commenting behavior on YouTube and Yahoo! News~\cite{siersdorfer2014analyzing}, or detect hoaxes, frauds, and deception in online documents~\cite{afroz2012detecting}. A large body of work targeted controversy in the political domain~\cite{mejova2014controversy,adamic2005political}, and studied controversy detection using social media network structure and content~\cite{garimella2016quantifying}.  

It has become clear that users polarization is a dominating aspect of online discussion~\cite{guess2018selective,ugander2012structural,guerra2013measure}. Indeed, users tend to confine their attention on a limited set of pages, thus determining a sharp community structure among news outlets \cite{schmidt2017anatomy}. Previous works~\cite{conti2017s} showed that it is difficult to carry out an automatic classification of misinformation considering only structural properties of content propagation cascades. Thus, to address misinformation problem properly, users' behavior has to be taken into account. 

In this work, we introduce a general framework for a timely identification of polarizing content on social media and, thus, the ``prediction'' of future fake news topics. Moreover, the output of our framework --i.e., whether a topic is susceptible to misinformation-- may be used as a new feature in a classifier aiming at recognizing fake news. The novelty of our contribution is three-fold:
\begin{enumerate}
\item To our knowledge, this is the first work dealing with the problem of the early detection of possible topics for fake news;
\item We introduce new features that account for how news are presented and perceived on the social network;
\item We provide a general framework that is easily extensible to different social media platforms. 
\end{enumerate}

\section{A framework for the early warning}\label{framework}
\begin{figure}[tb]
	\centering
	\includegraphics[width =.8\textwidth]{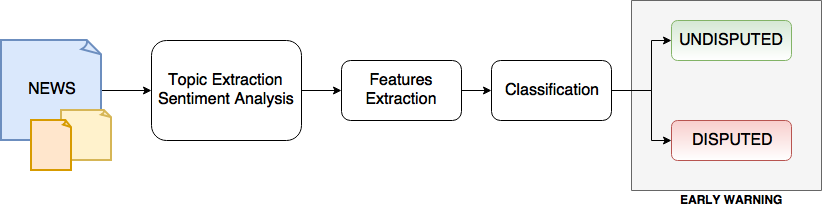}
	\caption{Overview of the proposed framework.}\label{fig:diagram}
\end{figure}

In this paper, we introduce a general framework to promptly determine polarizing content and detect potential breeding ground (\textit{early warning}) of either hoaxes, or fake, or unsubstantiated news. The proposed approach is suitable to different social media platforms --e.g., Facebook, Twitter-- and consists in four main phases, as shown in Figure~\ref{fig:diagram}:
\begin{description}
\item [1. Data collection:] First, we identify two categories of news sources: 1) official, and 2) fake --i.e., aiming at disseminating unsubstantiated or fake information. Then, for each category, we collect all data available on the platform under analysis.
\item [2. Topic extraction and sentiment analysis:] Second, we extract the topics (\textit{entities}) and sentiment associated to our data textual content. Entity extraction adds semantic knowledge to content to help understand the subject and context of the text that is being analyzed, allowing to identify items such as persons, places, and organizations that are present in the input text.
\item [3. Features definition:] We now use the information collected in the previous steps to derive a series of features that take into account how information is presented and perceived on the platform.
\item [4. Classification:] Finally, we perform the classification task using different state-of-the-art machine learning algorithms and comparing their results. Once detected the best algorithms (and the related feature sets), we are ready to classify entities and, thus, identify potential targets for fake news.
\end{description}

\subsection{Features}\label{sec:feat}
For the sake of simplicity, in the following we define our features for Facebook. However, the adaptation of the same features to other social media (e.g., Twitter, Instagram, YouTube) is straightforward. Let $e$ be an entity --i.e., one of the main items/topics in a Facebook post. We say a user to be engaged in entity $e$ if she/he left more than $95\%$ of her/his comments on posts containing $e$\footnote{Notice that a user may be engaged with more than one entity.}. We define the following features: 
\begin{enumerate}
\item The \textit{presentation distance} $d_p(e)$ i.e., the absolute difference between the maximum and the minimum value of the sentiment score of all posts containing entity $e$;
\item The \textit{mean response distance} $d_r(e)$ i.e., the absolute difference between the mean sentiment score on the posts containing the entity and the mean sentiment score on the related comments;\footnote{We also consider minimum, maximum, and the standard deviation for this measure.}
\item The \textit{controversy} of the entity $C(e)$:
	$$C(e)  = \begin{cases}  0, \mbox{ if } d_p(e) < \delta_p \\ 1, \mbox{ if } d_p(e) \geq \delta_p \end{cases}$$
	where $\delta_p$ is a specific threshold dependent on the data;
\item The \textit{perception} of the entity $P(e)$ as:
	$$P(e)  = \begin{cases}  0, \mbox{ if } d_r(e) < \delta_r \\ 1, \mbox{ if } d_r(e) \geq \delta_r \end{cases}$$
	where $\delta_r$ is a specific threshold dependent on the data;
\item The \textit{captivation} of the entity $\kappa(e)$:
	$$\kappa(e) =\begin{cases} 0, \mbox{ if } u_e  < \rho_e \\  1, \mbox{ if } u_e \geq \rho_e \end{cases}\rho_e\in[0,1]$$
	where $u_e$ is the fraction of users engaged in entity $e$ and $\rho_e$ is a threshold dependent on the data.
\end{enumerate}

Let $E$ be the set of all entities and $D$ the (number of) entities appearing in both categories --official and fake-- of news sources (hereafter referred as \textit{disputed entities}). To select the thresholds $\delta_p$, $\delta_r$, and $\rho_e$, for each considered feature $C(e)$, $P(e)$, and $\kappa(e)$, respectively, we define the following pairs:
\begin{itemize}
\item $(E_{\delta_p}, D_{\delta_p})$, where $E_{\delta_p}$ (respectively, $D_{\delta_p}$) is the number of all (respectively, disputed) entities in $E$ for which $d_p(e) \ge \delta_p$;
\item $(E_{\delta_r}, D_{\delta_r})$, where $E_{\delta_r}$ (respectively, $D_{\delta_r}$) is the number of all (respectively, disputed) entities in $E$ for which $d_r(e) \ge \delta_r$;
\item $(E_{\rho_e}, D_{\rho_e})$, where $E_{\rho_e}$ (respectively, $D_{\rho_e}$) is the number of all (respectively, disputed) entities in $E$ for which $u_e \ge \rho_e$.
\end{itemize}

A coherent and deep analysis of such metrics allows to determine the thresholds $\delta_p$, $\delta_r$, and $\rho_e$, that are clearly dependent on the data --and, thus, on the specific platform under analysis. In Section \ref{usecase} we exhaustively discuss a real use-case of our framework on Facebook. However, the adaptation to similar platforms is straightforward.

\subsection{Classification}\label{classification}
To identify topics that are potential targets for fake news, we compare the performance of several state-of-the-art classification algorithms, thus select the best ones and extract a set of features capable of ensuring a noteworthy level of accuracy. 
To this aim, we rely on the Python scikit-Learn package~\cite{scikit-learn} and, on the basis of the most recent literature~\cite{alsmadi2018term,ozel2017detection,khatua2017cricket,antonakaki2016exploiting,hemsley2017call,van2016social,vosoughi2016tweet,chang2017predicting}, we consider the following classifiers: Linear Regression (LIN)~\cite{rifkin2007notes}, Logistic Regression (LOG)~\cite{schmidt2017minimizing}, 
Support Vector Machine (SVM)~\cite{cortes1995support} through support vector classification,
K-Nearest Neighbors (KNN)~\cite{bentley1975multidimensional},
and Neural Network Models (NN)~\cite{rumelhart1988learning} through the Multi-layer Perceptron L-BFGS algorithm, and Decision Trees (DT)~\cite{breiman1984ra}.
To validate the results, we split the data into training ($60\%$) and test sets ($40\%$) and make use of the following metrics:
\begin{itemize}
\item $\mathit{Accuracy}$, that is the fraction of correctly classified examples (both true positives $T_p$ and true negatives $T_n$) among the total number of cases (N) examined:
$$\mathit{Accuracy} = \frac{T_p + T_n}{N}$$
\item $\mathit{Precision}$, that is the fraction of true positives ($T_p$) over the number of true positives plus the number of false positives ($F_p$):
$$\mathit{Precision} = \frac{T_p}{T_p + F_p}$$
\item $\mathit{Recall}$, that is the fraction of true positives ($T_p$) over the number of true positives plus the number of false negatives ($F_n$):
$$\mathit{Recall} = \frac{T_p}{T_p + F_n}$$
\item $\mathit{F_1}$\textit{-score}, that is the harmonic mean of $\mathit{Precision}$ and $\mathit{Recall}$~\cite{sokolova2009systematic}:
$$F_1\mbox{\textit{-score}} = 2 \cdot \frac{\mathit{Precision} \cdot \mathit{Recall}} {\mathit{Precision} + \mathit{Recall}} $$
\item Finally, the $\mathit{False\ Positive\ (FP)\ Rate}$ (or \textit{Inverse Recall}), that is the fraction of false positives ($F_p$) over the number of false positives plus the number of true negatives ($T_n$):
$$\mathit{FP\ Rate} = \frac {F_p}{F_p+T_n}$$
\end{itemize}
To evaluate and compare the classifiers output we measure the accuracy of the predicted values through the \textit{Area Under the ROC Curve} (AUC), where the \textit{Receiver Operating Characteristic} (ROC) is the curve that plots the \textit{Recall} against the \textit{FP Rate} at various thresholds settings~\cite{metz1978basic}.

\section{A Real Use-case: Facebook}\label{usecase}
In this section, we describe a real use-case of the proposed framework on Facebook. Our final aim is to identify the topics that are most likely to become a target for future fake news.

\subsection{Data collection}\label{dataset}
We identify two main categories of Facebook pages associated to: 
\begin{enumerate}
	\item Italian official newspapers (\textit{official});
	\item Italian websites that disseminate either hoaxes or unsubstantiated information or fake news (\textit{fake}).
\end{enumerate}
To produce our dataset, for set  \textit{(1)} we followed for the exhaustive list provided by ADS\footnote{ADS is an association for the verification of newspaper circulation in Italy. Their website provides an exhaustive list of Italian newspaper supplying documentation of their geographical and periodical diffusion.} \cite{ads}, while for set \textit{(2)} we relied on the lists provided by very active Italian debunking sites \cite{bufalenet, butac}. To validate the list, all pages have then been manually checked by looking at their self-description and the type of promoted content.
For each page, we downloaded all the posts in the period 31.07--12.12 2016, as well as all the related likes and comments. The exact breakdown of the dataset is provided in Table~\ref{tab:data}. The entire data collection process was performed exclusively by means of the Facebook Graph API \cite{fb_graph_api}, which is publicly available and can be used through one's personal Facebook user account. We used only public available data (users with privacy restrictions are not included in our dataset). Data was downloaded from Facebook pages that are public entities. 
When allowed by users' privacy specifications, we accessed public personal information. However, in our study we used fully anonymized and aggregated data. We abided by the terms, conditions, and privacy policies of Facebook.

\begin{table}[tb]
	\centering
	\caption{Breakdown of the dataset (Facebook).}
	\label{tab:data}
	\begin{tabular}{lcc}
		\toprule
		 & Official & Fake\\
		\midrule
		\textit{Pages} & $58$ &  $17$\\
		\textit{Posts} & $333,547$ & $51,535$ \\
		\textit{Likes} &  $74,822,459$ & $1,568,379$\\
		\textit{Comments} & $10,160,830$  & $505,821$ \\
		\textit{Shares} & $31,060,302$ & $2,730,476$ \\
		\bottomrule
	\end{tabular}
\end{table}

\subsection{Topic extraction and sentiment analysis}\label{sec:entities}
To perform topic extraction and sentiment analysis, we rely on Dandelion API~\cite{dandelion}, that is particularly suited for the Italian language and gets good performances on short texts as well \cite{canales2017evaluation}.
By means of the Dandelion API service, we extract the main entities and the sentiment score associated to each post of our dataset, whether it has a textual description or a link to an external document. Entities represent the main items in the text that, according to the service specifications, could fall in one of six categories: person, works, organizations, places, events, or concepts. Thus, for each post we get a list of entities and their related confidence level, and a sentiment score ranging from $-1.0$ (totally negative) to $1.0$ (absolutely positive). During the analysis, we only considered entities with a confidence level greater than or equal to $0.6$, hereafter referred as sample $E_1$. Moreover, we selected all entities with a confidence level greater than or equal to $0.9$ and occurring in at least 100 posts. For these entities, hereafter referred as sample $E_2$, we selected all the posts where they appeared and run Dandelion API to extract the sentiment score of the related comments. Details of both samples are shown in Table~\ref{tab:sample}.

\begin{table}[ht!]
\caption{Entities Samples.}
\label{tab:sample}
  \begin{tabular}{lcccc}
    \hline
    \multirow{2}{*}{} &
      \multicolumn{2}{c}{$E_1$} &
      \multicolumn{2}{c}{$E_2$} \\
    & Official & Fake & Official & Fake \\
    \hline
    \textit{Entities}& $82,589$ & $19,651$  & $1,170$ & $763$ \\
    \hline
    \textit{Posts} & $121,833$ & $5,995$& $16,098$ & $8,234$ \\
    \hline
    \textit{Comments} & $6,022,299$ & $135,988$ & $1,241,703$ & $171,062$ \\
    \hline
  \end{tabular}
\end{table}

\subsection{Features}
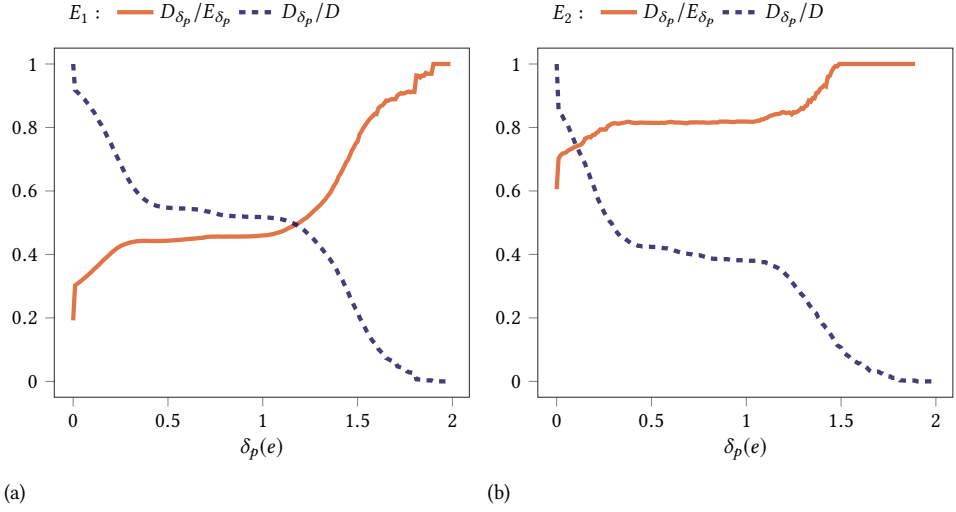
\begin{figure}[tb]
	\begin{subfigure}[]	
		\centering
		\resizebox{0.45\linewidth}{!}{
\begin{tikzpicture}

\definecolor{color0}{rgb}{0.85,0.325,0.098}
\definecolor{color1}{rgb}{0.101960784313725,0.0470588235294118,0.317647058823529}

\begin{axis}[
xlabel={$\delta_p(e)$},
xmin=-0.0995, xmax=2.0895,
ymin=-0.0498931352094041, ymax=1.04999491120045,
tick align=outside,
tick pos=left,
x grid style={white!80.0!black},
y grid style={white!80.0!black},
axis line style={darkgray!60.0!black},
legend columns=2,
legend entries={{$D_{\delta_p}/E_{\delta_p}$},{$D_{\delta_p}/D$}},
legend style={at={(0.13,1.05)}, anchor=south west, draw=none},
]
\addplot [line width=2.0pt, color0, opacity=0.8]
table {%
0 0.192367189488731
0.01 0.302409355941291
0.02 0.306941710177181
0.03 0.311086664221978
0.04 0.315873100887294
0.05 0.320654948581431
0.06 0.32581837184129
0.07 0.330803784765364
0.08 0.336194395453655
0.09 0.341619275365231
0.1 0.347245167685445
0.11 0.353371346206838
0.12 0.35849591280654
0.13 0.363980723971758
0.14 0.369820983667067
0.15 0.376111756753552
0.16 0.382329945269742
0.17 0.387400980761977
0.18 0.392775645679909
0.19 0.398669054173237
0.2 0.404260582737768
0.21 0.409611892873355
0.22 0.414094214538127
0.23 0.419021281712156
0.24 0.422875131164743
0.25 0.426339144215531
0.26 0.429249268768281
0.27 0.431388851935613
0.28 0.433763228638514
0.29 0.43584925357553
0.3 0.437692880706609
0.31 0.438558480482432
0.32 0.439958820501507
0.33 0.440962506994964
0.34 0.441913267236812
0.35 0.442217026332937
0.36 0.44244019509174
0.37 0.442680638098217
0.38 0.442722504830632
0.39 0.442816297355254
0.4 0.44266027561414
0.41 0.442339407103495
0.42 0.442219171444723
0.43 0.442207187132844
0.44 0.442103550656851
0.45 0.442287299127172
0.46 0.44241308793456
0.47 0.442648506727929
0.48 0.442864773801208
0.49 0.442898884727767
0.5 0.442970713020554
0.51 0.443856894984715
0.52 0.44408588804766
0.53 0.444605628548929
0.54 0.444877967790138
0.55 0.44536802294169
0.56 0.445748532900487
0.57 0.446292359634863
0.58 0.446823607980633
0.59 0.447580813783298
0.6 0.447952432794573
0.61 0.44874648301348
0.62 0.448883468606754
0.63 0.449506870100312
0.64 0.449985221466875
0.65 0.450602919399196
0.66 0.45129815034787
0.67 0.451914893617021
0.68 0.45228038738854
0.69 0.453028033383266
0.7 0.454941923226391
0.71 0.45520256232687
0.72 0.455947519332696
0.73 0.456090651558074
0.74 0.456052125240511
0.75 0.456064193633254
0.76 0.456482500550297
0.77 0.456166681380772
0.78 0.456145782652926
0.79 0.456043226006466
0.8 0.456043956043956
0.81 0.456337025668402
0.82 0.456366874443455
0.83 0.456388096935139
0.84 0.45627562845427
0.85 0.4561872909699
0.86 0.456392622694592
0.87 0.456337146943153
0.88 0.456370138578453
0.89 0.456410944206009
0.9 0.456488686164028
0.91 0.456552897966496
0.92 0.456688326237957
0.93 0.45695067264574
0.94 0.45714798528224
0.95 0.457547169811321
0.96 0.457700260721028
0.97 0.458207678804519
0.98 0.458637469586375
0.99 0.459333814768009
1 0.459730633643677
1.01 0.460157737285831
1.02 0.460679258990193
1.03 0.461419472247498
1.04 0.462511415525114
1.05 0.463557652663887
1.06 0.464952626253335
1.07 0.466845587895053
1.08 0.468553897643285
1.09 0.470395659900851
1.1 0.472220915463142
1.11 0.474705770690964
1.12 0.47780703854441
1.13 0.480818538048474
1.14 0.48400508706711
1.15 0.486871384067646
1.16 0.489239526639037
1.17 0.492662685962959
1.18 0.495822012610858
1.19 0.499375975039002
1.2 0.502901762161021
1.21 0.507751937984496
1.22 0.511804984935634
1.23 0.515244583426402
1.24 0.520507254655547
1.25 0.525928950592078
1.26 0.531898355754858
1.27 0.537552794270674
1.28 0.54298273155416
1.29 0.548719268889175
1.3 0.554210491499636
1.31 0.560019032082654
1.32 0.567569463934886
1.33 0.573376623376623
1.34 0.582112970711297
1.35 0.58961900354774
1.36 0.598333733878074
1.37 0.609860215953796
1.38 0.620476148949158
1.39 0.629592487920503
1.4 0.645600388915897
1.41 0.654288597376387
1.42 0.665606108813236
1.43 0.677568512461937
1.44 0.691160553217078
1.45 0.698529411764706
1.46 0.712728263832107
1.47 0.727776966866151
1.48 0.73689138576779
1.49 0.74937343358396
1.5 0.755380577427821
1.51 0.777864838393732
1.52 0.787135973182485
1.53 0.797780797101449
1.54 0.807034684904739
1.55 0.815208721084818
1.56 0.824036802760207
1.57 0.830059245400686
1.58 0.836290592567794
1.59 0.842652329749104
1.6 0.842027376988531
1.61 0.86095566078347
1.62 0.86657433056325
1.63 0.869306930693069
1.64 0.8752
1.65 0.884474327628362
1.66 0.883838383838384
1.67 0.887032085561497
1.68 0.889429763560501
1.69 0.888649425287356
1.7 0.889050036258158
1.71 0.900770712909441
1.72 0.905146316851665
1.73 0.908324552160169
1.74 0.906923950056754
1.75 0.908224076281287
1.76 0.911465892597968
1.77 0.913183279742765
1.78 0.910891089108911
1.79 0.912820512820513
1.8 0.912220309810671
1.81 0.963768115942029
1.82 0.962962962962963
1.83 0.957627118644068
1.84 0.963302752293578
1.85 0.962962962962963
1.86 0.971830985915493
1.87 0.969230769230769
1.88 0.969230769230769
1.89 0.969230769230769
1.9 1
1.91 1
1.92 1
1.93 1
1.94 1
1.95 1
1.96 1
1.97 1
1.98 1
1.99 1
};
\addplot [line width=2.0pt, color1, opacity=0.8, dashed]
table {%
0 0.999847336013434
0.01 0.918477431173986
0.02 0.912421759706885
0.03 0.906569640221872
0.04 0.900361304768205
0.05 0.894916289247367
0.06 0.888402625820569
0.07 0.880667650501247
0.08 0.873034451172968
0.09 0.865095923871559
0.1 0.855681644700015
0.11 0.848353773344868
0.12 0.83690397435245
0.13 0.826370159279426
0.14 0.81578545621088
0.15 0.806981832985599
0.16 0.796295353926009
0.17 0.783929571014198
0.18 0.770800468169559
0.19 0.759096229199532
0.2 0.748409750139942
0.21 0.734720879344563
0.22 0.721540888504402
0.23 0.709378657574678
0.24 0.697267314640476
0.25 0.68449442776449
0.26 0.672128644852679
0.27 0.660068189913999
0.28 0.648669278917103
0.29 0.637372143911251
0.3 0.627906976744186
0.31 0.618034705612946
0.32 0.608925754414534
0.33 0.601496107068343
0.34 0.594270011704239
0.35 0.587094804335657
0.36 0.581649788814819
0.37 0.576153885298458
0.38 0.571319525723882
0.39 0.567452038064221
0.4 0.563940766373213
0.41 0.559615286753855
0.42 0.557325326955371
0.43 0.555442471121062
0.44 0.553152511322579
0.45 0.551829423439011
0.46 0.550455447559921
0.47 0.549081471680831
0.48 0.548470815734568
0.49 0.547656607806218
0.5 0.547249503842044
0.51 0.546740623886825
0.52 0.546231743931607
0.53 0.545875527962954
0.54 0.545417536003257
0.55 0.545315760012213
0.56 0.545010432039082
0.57 0.544857768052516
0.58 0.544755992061473
0.59 0.544654216070429
0.6 0.54439977609282
0.61 0.543789120146557
0.62 0.543178464200295
0.63 0.542720472240598
0.64 0.542313368276424
0.65 0.541957152307771
0.66 0.541346496361508
0.67 0.540430512442115
0.68 0.5394636405272
0.69 0.53864943259885
0.7 0.536155920818279
0.71 0.535189048903364
0.72 0.534069513001883
0.73 0.532542873136227
0.74 0.53071090529744
0.75 0.529286041422828
0.76 0.527657625566129
0.77 0.525876545722864
0.78 0.524807897816905
0.79 0.523993689888555
0.8 0.523739249910946
0.81 0.522009058063203
0.82 0.521601954099028
0.83 0.521347514121419
0.84 0.520940410157244
0.85 0.520584194188591
0.86 0.520075314233372
0.87 0.519617322273676
0.88 0.519515546282632
0.89 0.519515546282632
0.9 0.51946465828711
0.91 0.518701338354282
0.92 0.518599562363239
0.93 0.518548674367717
0.94 0.518446898376673
0.95 0.518294234390107
0.96 0.518141570403542
0.97 0.518039794412498
0.98 0.517988906416976
0.99 0.517887130425932
1 0.517632690448323
1.01 0.516614930537886
1.02 0.516309602564755
1.03 0.516106050582668
1.04 0.515444506640883
1.05 0.514935626685665
1.06 0.514426746730446
1.07 0.513408986820009
1.08 0.512950994860312
1.09 0.511831458958832
1.1 0.510813699048394
1.11 0.50903261920513
1.12 0.507811307312605
1.13 0.505775787491731
1.14 0.503536715688769
1.15 0.501043203908198
1.16 0.498600580123149
1.17 0.495445524400794
1.18 0.492188692687395
1.19 0.488677420996387
1.2 0.485064373314335
1.21 0.479975573762149
1.22 0.475446542160704
1.23 0.469543534680169
1.24 0.463691415195155
1.25 0.458806167625057
1.26 0.452699608162434
1.27 0.446898376672943
1.28 0.440028497277492
1.29 0.433871049819348
1.3 0.426339626482113
1.31 0.419266195104575
1.32 0.411632995776296
1.33 0.404406900412193
1.34 0.396468373110783
1.35 0.389038725764592
1.36 0.380082438552745
1.37 0.370769935372246
1.38 0.362068088138008
1.39 0.35143249707394
1.4 0.337896290265126
1.41 0.329957762963717
1.42 0.319373059895171
1.43 0.305735077095313
1.44 0.292453310264109
1.45 0.280392855325429
1.46 0.266144216579309
1.47 0.253727545671976
1.48 0.240293114854206
1.49 0.228232659915526
1.5 0.219683476667854
1.51 0.202076230217292
1.52 0.191186199175614
1.53 0.1792784082235
1.54 0.168133937204214
1.55 0.156022594270012
1.56 0.14584499516564
1.57 0.135463844079182
1.58 0.127118212813597
1.59 0.119637677471884
1.6 0.115821077807745
1.61 0.101775991043713
1.62 0.0955167675945244
1.63 0.0893593201363798
1.64 0.0835072006513663
1.65 0.0736349295201262
1.66 0.071243193730599
1.67 0.0675283700575034
1.68 0.0650857462724543
1.69 0.0629484504605364
1.7 0.0623886825097959
1.71 0.0475802758129357
1.72 0.0456465319831052
1.73 0.0438654521398402
1.74 0.0406595084219633
1.75 0.0387766525876546
1.76 0.0319576611877258
1.77 0.0289043814564144
1.78 0.0280901735280647
1.79 0.0271741896086713
1.8 0.0269706376265839
1.81 0.0067681034044069
1.82 0.00661543941784133
1.83 0.00575034349396977
1.84 0.00534323952979492
1.85 0.00529235153427306
1.86 0.00351127169100809
1.87 0.00320594371787695
1.88 0.00320594371787695
1.89 0.00320594371787695
1.9 0.000152663986565569
1.91 0.000152663986565569
1.92 0.000152663986565569
1.93 0.000152663986565569
1.94 0.000152663986565569
1.95 0.000152663986565569
1.96 0.000101775991043713
1.97 0.000101775991043713
1.98 0.000101775991043713
1.99 0.000101775991043713
};
\end{axis}
\node at (0.5,6.25) {$E_1:$};
\end{tikzpicture}}
	\end{subfigure}
	\begin{subfigure}[]	
		\centering
		\resizebox{0.45\linewidth}{!}{
\begin{tikzpicture}

\definecolor{color0}{rgb}{0.85,0.325,0.098}
\definecolor{color1}{rgb}{0.101960784313725,0.0470588235294118,0.317647058823529}

\begin{axis}[
xlabel={$\delta_p(e)$},
xmin=-0.0995, xmax=2.0895,
ymin=-0.05, ymax=1.05,
tick align=outside,
tick pos=left,
x grid style={white!80.0!black},
y grid style={white!80.0!black},
axis line style={darkgray!60.0!black},
legend columns=2,
legend entries={{$D_{\delta_p}/E_{\delta_p}$},{$D_{\delta_p}/D$}},
legend style={at={(0.13, 1.05)}, anchor=south west, draw=none},
]
\addplot [line width=2.0pt, color0, opacity=0.8]
table {%
0 0.605480868665977
0.01 0.702247191011236
0.02 0.711815561959654
0.03 0.717741935483871
0.04 0.719076005961252
0.05 0.721633888048411
0.06 0.727062451811874
0.07 0.730647709320695
0.08 0.732689210950081
0.09 0.737190082644628
0.1 0.739424703891709
0.11 0.742463393626184
0.12 0.743838028169014
0.13 0.746858168761221
0.14 0.752539242843952
0.15 0.764761904761905
0.16 0.767716535433071
0.17 0.769849246231156
0.18 0.768833849329205
0.19 0.776357827476038
0.2 0.777049180327869
0.21 0.78310502283105
0.22 0.787058823529412
0.23 0.793939393939394
0.24 0.794044665012407
0.25 0.792884371029225
0.26 0.793774319066148
0.27 0.804521276595745
0.28 0.807113543091655
0.29 0.811977715877437
0.3 0.812857142857143
0.31 0.813411078717201
0.32 0.811851851851852
0.33 0.81203007518797
0.34 0.813455657492355
0.35 0.814528593508501
0.36 0.815503875968992
0.37 0.817896389324961
0.38 0.817749603803486
0.39 0.81629392971246
0.4 0.815112540192926
0.41 0.814516129032258
0.42 0.813915857605178
0.43 0.814332247557003
0.44 0.815660685154976
0.45 0.81505728314239
0.46 0.81505728314239
0.47 0.814754098360656
0.48 0.814754098360656
0.49 0.814754098360656
0.5 0.814449917898194
0.51 0.814449917898194
0.52 0.814449917898194
0.53 0.814449917898194
0.54 0.814144736842105
0.55 0.814876033057851
0.56 0.814569536423841
0.57 0.814262023217247
0.58 0.815614617940199
0.59 0.816360601001669
0.6 0.817725752508361
0.61 0.816806722689076
0.62 0.815878378378378
0.63 0.815254237288136
0.64 0.815699658703072
0.65 0.815384615384615
0.66 0.8147512864494
0.67 0.814432989690722
0.68 0.813471502590674
0.69 0.813471502590674
0.7 0.815972222222222
0.71 0.815652173913043
0.72 0.816112084063047
0.73 0.816112084063047
0.74 0.816112084063047
0.75 0.816112084063047
0.76 0.815465729349736
0.77 0.816254416961131
0.78 0.815602836879433
0.79 0.814946619217082
0.8 0.814946619217082
0.81 0.81508078994614
0.82 0.816216216216216
0.83 0.815884476534296
0.84 0.815884476534296
0.85 0.815884476534296
0.86 0.818511796733212
0.87 0.818511796733212
0.88 0.818511796733212
0.89 0.818511796733212
0.9 0.818511796733212
0.91 0.817850637522769
0.92 0.817184643510055
0.93 0.817184643510055
0.94 0.817184643510055
0.95 0.818681318681319
0.96 0.818681318681319
0.97 0.818681318681319
0.98 0.818681318681319
0.99 0.818681318681319
1 0.818681318681319
1.01 0.818014705882353
1.02 0.818014705882353
1.03 0.818014705882353
1.04 0.818014705882353
1.05 0.819521178637201
1.06 0.821033210332103
1.07 0.821561338289963
1.08 0.825515947467167
1.09 0.825515947467167
1.1 0.828625235404896
1.11 0.83047619047619
1.12 0.834615384615385
1.13 0.837573385518591
1.14 0.838582677165354
1.15 0.841269841269841
1.16 0.842
1.17 0.842424242424242
1.18 0.844262295081967
1.19 0.848547717842324
1.2 0.847780126849894
1.21 0.845161290322581
1.22 0.846827133479212
1.23 0.847006651884701
1.24 0.841379310344828
1.25 0.848699763593381
1.26 0.847087378640777
1.27 0.85
1.28 0.854219948849105
1.29 0.86096256684492
1.3 0.858695652173913
1.31 0.866477272727273
1.32 0.869186046511628
1.33 0.884735202492212
1.34 0.881410256410256
1.35 0.89261744966443
1.36 0.89198606271777
1.37 0.909090909090909
1.38 0.910505836575875
1.39 0.922131147540984
1.4 0.926086956521739
1.41 0.932432432432432
1.42 0.929245283018868
1.43 0.95959595959596
1.44 0.962162162162162
1.45 0.976744186046512
1.46 0.987341772151899
1.47 0.993150684931507
1.48 0.992537313432836
1.49 1
1.5 1
1.51 1
1.52 1
1.53 1
1.54 1
1.55 1
1.56 1
1.57 1
1.58 1
1.59 1
1.6 1
1.61 1
1.62 1
1.63 1
1.64 1
1.65 1
1.66 1
1.67 1
1.68 1
1.69 1
1.7 1
1.71 1
1.72 1
1.73 1
1.74 1
1.75 1
1.76 1
1.77 1
1.78 1
1.79 1
1.8 1
1.81 1
1.82 1
1.83 1
1.84 1
1.85 1
1.86 1
1.87 1
1.88 1
1.89 1
1.9 nan
1.91 nan
1.92 nan
1.93 nan
1.94 nan
1.95 nan
1.96 nan
1.97 nan
1.98 nan
1.99 nan
};
\addplot [line width=2.0pt, color1, opacity=0.8, dashed]
table {%
0 1
0.01 0.85397096498719
0.02 0.843723313407344
0.03 0.836037574722459
0.04 0.824081981212639
0.05 0.81468830059778
0.06 0.805294619982921
0.07 0.789923142613151
0.08 0.777113578138343
0.09 0.761742100768574
0.1 0.746370623398804
0.11 0.736122971818958
0.12 0.721605465414176
0.13 0.710503842869342
0.14 0.69598633646456
0.15 0.685738684884714
0.16 0.666097352690009
0.17 0.654141759180188
0.18 0.636208368915457
0.19 0.622544833475662
0.2 0.607173356105892
0.21 0.585824081981213
0.22 0.57130657557643
0.23 0.55935098206661
0.24 0.546541417591802
0.25 0.532877882152007
0.26 0.522630230572161
0.27 0.51665243381725
0.28 0.503842869342442
0.29 0.497865072587532
0.3 0.485909479077711
0.31 0.476515798462852
0.32 0.46797608881298
0.33 0.461144321093083
0.34 0.454312553373185
0.35 0.450042698548249
0.36 0.449188727583262
0.37 0.444918872758326
0.38 0.44064901793339
0.39 0.436379163108454
0.4 0.432963279248506
0.41 0.431255337318531
0.42 0.429547395388557
0.43 0.426985482493595
0.44 0.426985482493595
0.45 0.425277540563621
0.46 0.425277540563621
0.47 0.424423569598634
0.48 0.424423569598634
0.49 0.424423569598634
0.5 0.423569598633646
0.51 0.423569598633646
0.52 0.423569598633646
0.53 0.423569598633646
0.54 0.422715627668659
0.55 0.421007685738685
0.56 0.420153714773698
0.57 0.41929974380871
0.58 0.41929974380871
0.59 0.417591801878736
0.6 0.417591801878736
0.61 0.415029888983775
0.62 0.412467976088813
0.63 0.410760034158839
0.64 0.408198121263877
0.65 0.40734415029889
0.66 0.405636208368915
0.67 0.404782237403928
0.68 0.402220324508967
0.69 0.402220324508967
0.7 0.401366353543979
0.71 0.400512382578992
0.72 0.397950469684031
0.73 0.397950469684031
0.74 0.397950469684031
0.75 0.397950469684031
0.76 0.396242527754056
0.77 0.394534585824082
0.78 0.392826643894108
0.79 0.391118701964133
0.8 0.391118701964133
0.81 0.387702818104184
0.82 0.386848847139197
0.83 0.38599487617421
0.84 0.38599487617421
0.85 0.38599487617421
0.86 0.385140905209223
0.87 0.385140905209223
0.88 0.385140905209223
0.89 0.385140905209223
0.9 0.385140905209223
0.91 0.383432963279249
0.92 0.381725021349274
0.93 0.381725021349274
0.94 0.381725021349274
0.95 0.381725021349274
0.96 0.381725021349274
0.97 0.381725021349274
0.98 0.381725021349274
0.99 0.381725021349274
1 0.381725021349274
1.01 0.3800170794193
1.02 0.3800170794193
1.03 0.3800170794193
1.04 0.3800170794193
1.05 0.3800170794193
1.06 0.3800170794193
1.07 0.377455166524338
1.08 0.375747224594364
1.09 0.375747224594364
1.1 0.375747224594364
1.11 0.372331340734415
1.12 0.370623398804441
1.13 0.365499573014518
1.14 0.363791631084543
1.15 0.362083689154569
1.16 0.359521776259607
1.17 0.356105892399658
1.18 0.351836037574722
1.19 0.349274124679761
1.2 0.342442356959863
1.21 0.335610589239966
1.22 0.330486763450043
1.23 0.326216908625107
1.24 0.312553373185312
1.25 0.306575576430401
1.26 0.298035866780529
1.27 0.290350128095645
1.28 0.285226302305722
1.29 0.274978650725875
1.3 0.269854824935952
1.31 0.260461144321093
1.32 0.25533731853117
1.33 0.242527754056362
1.34 0.234842015371477
1.35 0.227156276686593
1.36 0.218616567036721
1.37 0.213492741246798
1.38 0.199829205807003
1.39 0.192143467122118
1.4 0.181895815542272
1.41 0.176771989752348
1.42 0.168232280102477
1.43 0.162254483347566
1.44 0.15200683176772
1.45 0.143467122117848
1.46 0.133219470538002
1.47 0.123825789923143
1.48 0.113578138343296
1.49 0.110162254483348
1.5 0.105892399658412
1.51 0.0930828351836038
1.52 0.0879590093936806
1.53 0.0828351836037575
1.54 0.0794192997438087
1.55 0.071733561058924
1.56 0.0666097352690009
1.57 0.0640478223740393
1.58 0.0606319385140905
1.59 0.0572160546541418
1.6 0.0563620836891546
1.61 0.0486763450042699
1.62 0.0478223740392827
1.63 0.0452604611443211
1.64 0.0409906063193851
1.65 0.0341588385994876
1.66 0.0307429547395389
1.67 0.0307429547395389
1.68 0.0307429547395389
1.69 0.0307429547395389
1.7 0.0298889837745517
1.71 0.0256191289496157
1.72 0.0247651579846285
1.73 0.0204953031596926
1.74 0.0162254483347566
1.75 0.0153714773697694
1.76 0.0153714773697694
1.77 0.0111016225448335
1.78 0.0111016225448335
1.79 0.0111016225448335
1.8 0.0111016225448335
1.81 0.00341588385994876
1.82 0.00341588385994876
1.83 0.00256191289496157
1.84 0.00256191289496157
1.85 0.00256191289496157
1.86 0.00256191289496157
1.87 0.00256191289496157
1.88 0.00256191289496157
1.89 0.00256191289496157
1.9 0
1.91 0
1.92 0
1.93 0
1.94 0
1.95 0
1.96 0
1.97 0
1.98 0
1.99 0
};
\end{axis}
\node at (0.5,6.25) {$E_2:$};
\end{tikzpicture}}
	\end{subfigure}
	\caption{We show in dashed blue $D_{\delta_p}/D$, the fraction of disputed entities with presentation distance greater than or equal to $\delta_p$, and in solid orange $D_{\delta_p}/E_{\delta_p}$, the ratio of disputed entities w.r.t. all entities with presentation distance greater than or equal to $\delta_p$, both for sample $E1$ (a) and $E2$ (b). The plateaus in the curves indicate that entities are well separated into a set on uncontroversial (low $\delta_p$) and a set of controversial entities (high $\delta_p$).}
	\label{fig:pres_dist}
\end{figure}

Following the features presented in Section \ref{sec:feat}, it is straightforward to compute the presentation distance $d_p(e)$ and the mean response distance $d_r(e)$ for each entity of our samples $E_1$ and $E_2$. To calculate controversy, perception and captivation, we first need to find proper thresholds for our data. To develop an intuition on how to determine such quantities, let us analyze the behavior of the number of disputed entities as a function of the thresholds. In Figure~\ref{fig:pres_dist} we show for both samples $E_1$ (left) and $E_2$ (right) the number of disputed entities $D_{\delta_p}$ with presentation distance $\geq \delta_p$ normalized with respect to the total number of disputed entities $D$. We observe that $D_{\delta_p}/D$ presents a plateau between two regions of monotonic decrease with respect to $\delta_p$. 
This behavior indicates that entities are clearly separated in two sets and that all the entities with $d_p(e)\geq \delta_p$ are indeed controversial. Consequently, we may take the inflection point corresponding to the second change in the curve concavity as our threshold $\delta_p(e)$, since it accounts for the majority of disputed entities. To do that, we fit our data to polynomial functions and compute all inflection points. We get the following thresholds for $C(e)$: $\delta_p(e) = 1.1$ for sample $E_1$ and $\delta_p(e) = 0.98$ for sample $E_2$.
Notice that the height of the plateau corresponds to the size of controversial news (i.e. $d_p(e)>\delta_p$): hence, for $E_1$ and $E_2$ we have that respectively $\sim55\%$ and $\sim40\%$ of the disputed entities are controversial.

In Figure~\ref{fig:pres_dist} we show the ratio $D_{\delta_p}/E_{\delta_p}$, that measures the correlation between disputed entities $D$ and all entities $E$ by varying presentation distances. Notice that the fact that $D_{\delta_p}/E_{\delta_p}\to 1$ when $\delta_p$ grows indicates that the all highly controversial entities are also disputed entities. Like $D_{\delta_p}/D$, also $D_{\delta_p}/E_{\delta_p}$ shows a plateau in the same region of $\delta_p$'s. We observe that the main difference between the two samples $E_1$ and $E_2$ is in the initial value of $D_{\delta_p}/E_{\delta_p}$ and the height of the plateaus. For sample $E_2$, the total number $D_{\delta_p =0}$ of disputed entities is about $\sim60\%$ of the total number of entities $E_{\delta_p =0}$, while for $E_1$ it is only $\sim20\%$. On the same footing, we have that the different height of the plateaus indicates that while for set $E_1$ $45\%$ of controversial news are also disputed (i.e. they belong both to the official and fake categories), for set $E2$ this ratio becomes $\sim80\%$

An analogous approach may be used to find the thresholds $\delta_r$ and $\rho_e$ for features perception $P(e)$ and captivation $\kappa(e)$, respectively. Notice that such features are only applicable to sample $E_2$, for which the sentiment score is available for comments too. 
 
\begin{figure}[h!]	
	\begin{subfigure}[]	
		\centering
		\resizebox{0.45\linewidth}{!}{
\begin{tikzpicture}

\definecolor{color0}{rgb}{0.929,0.694,0.125}
\definecolor{color1}{rgb}{0.101960784313725,0.0470588235294118,0.317647058823529}

\begin{axis}[
xlabel={$\delta_r(e)$},
xmin=-0.0995, xmax=2.0895,
ymin=-0.05, ymax=1.05,
tick align=outside,
tick pos=left,
x grid style={white!80.0!black},
y grid style={white!80.0!black},
axis line style={darkgray!60.0!black},
legend cell align={left},
legend style={at={(0.13, 1.05)}, anchor=south west, draw=none},
legend columns=2,
legend entries={{$D_{\delta_r}/E_{\delta_r}$},{$D_{\delta_r}/D$}}
]
\addplot [line width=2.0pt, color0, opacity=0.8]
table {%
0 0.605794102431454
0.01 0.606233766233766
0.02 0.606975533576262
0.03 0.607310704960836
0.04 0.607442348008386
0.05 0.6120826709062
0.06 0.613418530351438
0.07 0.614354579539368
0.08 0.617170626349892
0.09 0.618814573137575
0.1 0.621237000547345
0.11 0.621546961325967
0.12 0.621591541457986
0.13 0.624228827818284
0.14 0.625282167042889
0.15 0.626916524701874
0.16 0.626499143346659
0.17 0.627947096032202
0.18 0.631395348837209
0.19 0.635079458505003
0.2 0.638790035587189
0.21 0.64033512866547
0.22 0.644201578627808
0.23 0.646229307173513
0.24 0.650931677018634
0.25 0.655324511657215
0.26 0.659424920127796
0.27 0.666883116883117
0.28 0.670184696569921
0.29 0.669112741827885
0.3 0.678058783321941
0.31 0.682229965156794
0.32 0.679459843638948
0.33 0.683979517190929
0.34 0.67741935483871
0.35 0.672868217054264
0.36 0.676734693877551
0.37 0.674914675767918
0.38 0.645390070921986
0.39 0.641491395793499
0.4 0.629817444219067
0.41 0.600425079702444
0.42 0.58266818700114
0.43 0.560736196319018
0.44 0.544854881266491
0.45 0.497198879551821
0.46 0.457831325301205
0.47 0.405063291139241
0.48 0.382154882154882
0.49 0.35948905109489
0.5 0.340816326530612
0.51 0.305309734513274
0.52 0.274004683840749
0.53 0.243107769423559
0.54 0.221621621621622
0.55 0.202898550724638
0.56 0.184049079754601
0.57 0.18213058419244
0.58 0.164750957854406
0.59 0.148760330578512
0.6 0.13963963963964
0.61 0.14367816091954
0.62 0.123376623376623
0.63 0.123188405797101
0.64 0.115384615384615
0.65 0.104
0.66 0.11304347826087
0.67 0.11214953271028
0.68 0.11340206185567
0.69 0.121951219512195
0.7 0.123287671232877
0.71 0.126984126984127
0.72 0.103448275862069
0.73 0.111111111111111
0.74 0.1
0.75 0.102040816326531
0.76 0.0975609756097561
0.77 0.102564102564103
0.78 0.105263157894737
0.79 0.117647058823529
0.8 0.0689655172413793
0.81 0.0357142857142857
0.82 0.037037037037037
0.83 0.0416666666666667
0.84 0.0454545454545455
0.85 0.0454545454545455
0.86 0.05
0.87 0.05
0.88 0
0.89 0
0.9 0
0.91 0
0.92 0
0.93 0
0.94 0
0.95 0
0.96 0
0.97 0
0.98 0
0.99 0
1 0
1.01 0
1.02 0
1.03 0
1.04 0
1.05 0
1.06 0
1.07 0
1.08 0
1.09 0
1.1 0
1.11 0
1.12 0
1.13 0
1.14 0
1.15 0
1.16 0
1.17 0
1.18 0
1.19 0
1.2 0
1.21 0
1.22 0
1.23 0
1.24 0
1.25 0
1.26 0
1.27 0
1.28 0
1.29 0
1.3 0
1.31 0
1.32 0
1.33 0
1.34 0
1.35 0
1.36 0
1.37 0
1.38 0
1.39 0
1.4 nan
1.41 nan
1.42 nan
1.43 nan
1.44 nan
1.45 nan
1.46 nan
1.47 nan
1.48 nan
1.49 nan
1.5 nan
1.51 nan
1.52 nan
1.53 nan
1.54 nan
1.55 nan
1.56 nan
1.57 nan
1.58 nan
1.59 nan
1.6 nan
1.61 nan
1.62 nan
1.63 nan
1.64 nan
1.65 nan
1.66 nan
1.67 nan
1.68 nan
1.69 nan
1.7 nan
1.71 nan
1.72 nan
1.73 nan
1.74 nan
1.75 nan
1.76 nan
1.77 nan
1.78 nan
1.79 nan
1.8 nan
1.81 nan
1.82 nan
1.83 nan
1.84 nan
1.85 nan
1.86 nan
1.87 nan
1.88 nan
1.89 nan
1.9 nan
1.91 nan
1.92 nan
1.93 nan
1.94 nan
1.95 nan
1.96 nan
1.97 nan
1.98 nan
1.99 nan
};
\addplot [line width=2.0pt, color1, opacity=0.8, dashed]
table {%
0 1
0.01 0.996584116140051
0.02 0.995730145175064
0.03 0.993168232280102
0.04 0.989752348420154
0.05 0.986336464560205
0.06 0.983774551665243
0.07 0.979504696840307
0.08 0.976088812980359
0.09 0.971818958155423
0.1 0.969257045260461
0.11 0.960717335610589
0.12 0.953885567890692
0.13 0.950469684030743
0.14 0.946199829205807
0.15 0.942783945345858
0.16 0.936806148590948
0.17 0.932536293766012
0.18 0.927412467976089
0.19 0.921434671221179
0.2 0.919726729291204
0.21 0.913748932536294
0.22 0.906063193851409
0.23 0.900085397096499
0.24 0.894961571306576
0.25 0.888129803586678
0.26 0.881298035866781
0.27 0.877028181041845
0.28 0.867634500426986
0.29 0.856532877882152
0.3 0.847139197267293
0.31 0.836037574722459
0.32 0.816396242527754
0.33 0.798462852263023
0.34 0.771135781383433
0.35 0.741246797608881
0.36 0.707941929974381
0.37 0.675491033304868
0.38 0.621690862510675
0.39 0.573014517506405
0.4 0.530315969257045
0.41 0.482493595217763
0.42 0.436379163108454
0.43 0.390264730999146
0.44 0.35269000853971
0.45 0.303159692570453
0.46 0.259607173356106
0.47 0.218616567036721
0.48 0.193851409052092
0.49 0.168232280102477
0.5 0.142613151152861
0.51 0.117847993168232
0.52 0.0999146029035013
0.53 0.0828351836037575
0.54 0.0700256191289496
0.55 0.0597779675491033
0.56 0.0512382578992314
0.57 0.0452604611443211
0.58 0.0367207514944492
0.59 0.0307429547395389
0.6 0.0264730999146029
0.61 0.0213492741246798
0.62 0.0162254483347566
0.63 0.0145175064047822
0.64 0.0128095644748079
0.65 0.0111016225448335
0.66 0.0111016225448335
0.67 0.0102476515798463
0.68 0.00939368061485909
0.69 0.0085397096498719
0.7 0.00768573868488471
0.71 0.00683176771989752
0.72 0.00512382578992314
0.73 0.00512382578992314
0.74 0.00426985482493595
0.75 0.00426985482493595
0.76 0.00341588385994876
0.77 0.00341588385994876
0.78 0.00341588385994876
0.79 0.00341588385994876
0.8 0.00170794192997438
0.81 0.00085397096498719
0.82 0.00085397096498719
0.83 0.00085397096498719
0.84 0.00085397096498719
0.85 0.00085397096498719
0.86 0.00085397096498719
0.87 0.00085397096498719
0.88 0
0.89 0
0.9 0
0.91 0
0.92 0
0.93 0
0.94 0
0.95 0
0.96 0
0.97 0
0.98 0
0.99 0
1 0
1.01 0
1.02 0
1.03 0
1.04 0
1.05 0
1.06 0
1.07 0
1.08 0
1.09 0
1.1 0
1.11 0
1.12 0
1.13 0
1.14 0
1.15 0
1.16 0
1.17 0
1.18 0
1.19 0
1.2 0
1.21 0
1.22 0
1.23 0
1.24 0
1.25 0
1.26 0
1.27 0
1.28 0
1.29 0
1.3 0
1.31 0
1.32 0
1.33 0
1.34 0
1.35 0
1.36 0
1.37 0
1.38 0
1.39 0
1.4 0
1.41 0
1.42 0
1.43 0
1.44 0
1.45 0
1.46 0
1.47 0
1.48 0
1.49 0
1.5 0
1.51 0
1.52 0
1.53 0
1.54 0
1.55 0
1.56 0
1.57 0
1.58 0
1.59 0
1.6 0
1.61 0
1.62 0
1.63 0
1.64 0
1.65 0
1.66 0
1.67 0
1.68 0
1.69 0
1.7 0
1.71 0
1.72 0
1.73 0
1.74 0
1.75 0
1.76 0
1.77 0
1.78 0
1.79 0
1.8 0
1.81 0
1.82 0
1.83 0
1.84 0
1.85 0
1.86 0
1.87 0
1.88 0
1.89 0
1.9 0
1.91 0
1.92 0
1.93 0
1.94 0
1.95 0
1.96 0
1.97 0
1.98 0
1.99 0
};
\end{axis}
\node at (0.5,6.25) {$E_2:$};

\end{tikzpicture}}
	\end{subfigure}
	\begin{subfigure}[]	
		\centering
		\resizebox{0.45\linewidth}{!}{
\begin{tikzpicture}

\definecolor{color0}{rgb}{0.466,0.674,0.188}
\definecolor{color1}{rgb}{0.101960784313725,0.0470588235294118,0.317647058823529}

\begin{axis}[
xlabel={$\rho_e$},
xmin=-0.0495, xmax=1.0395,
ymin=0.0477369769427839, ymax=1.04534585824082,
tick align=outside,
tick pos=left,
x grid style={white!80.0!black},
y grid style={white!80.0!black},
axis line style={darkgray!60.0!black},
legend columns=2,
legend style={at={(0.13, 1.05)}, anchor=south west, draw=none},
legend entries={{$D_{\rho_e}/E_{\rho_e}$},{$D_{\rho_e}/D$}},
]
\addplot [line width=2.0pt, color0, opacity=0.8]
table {%
0 0.605480868665977
0.01 0.602110817941953
0.02 0.613523914238593
0.03 0.615950226244344
0.04 0.629741119807345
0.05 0.638993710691824
0.06 0.649797570850202
0.07 0.655244755244755
0.08 0.662832494608195
0.09 0.674542682926829
0.1 0.682946357085669
0.11 0.698466780238501
0.12 0.708746618575293
0.13 0.716981132075472
0.14 0.716981132075472
0.15 0.723061430010071
0.16 0.723061430010071
0.17 0.751451800232288
0.18 0.751451800232288
0.19 0.752325581395349
0.2 0.752325581395349
0.21 0.773030707610147
0.22 0.773030707610147
0.23 0.773030707610147
0.24 0.773030707610147
0.25 0.773030707610147
0.26 0.819148936170213
0.27 0.819148936170213
0.28 0.819148936170213
0.29 0.819148936170213
0.3 0.819148936170213
0.31 0.819148936170213
0.32 0.819148936170213
0.33 0.819148936170213
0.34 0.857954545454545
0.35 0.857954545454545
0.36 0.857954545454545
0.37 0.857954545454545
0.38 0.857954545454545
0.39 0.857954545454545
0.4 0.857954545454545
0.41 0.857549857549858
0.42 0.857549857549858
0.43 0.857549857549858
0.44 0.857549857549858
0.45 0.857549857549858
0.46 0.857549857549858
0.47 0.857549857549858
0.48 0.857549857549858
0.49 0.857549857549858
0.5 0.857549857549858
0.51 0.915966386554622
0.52 0.915966386554622
0.53 0.915966386554622
0.54 0.915966386554622
0.55 0.915966386554622
0.56 0.915966386554622
0.57 0.915966386554622
0.58 0.915966386554622
0.59 0.915966386554622
0.6 0.915966386554622
0.61 0.915966386554622
0.62 0.915966386554622
0.63 0.915966386554622
0.64 0.915966386554622
0.65 0.915966386554622
0.66 0.915966386554622
0.67 0.915966386554622
0.68 0.915966386554622
0.69 0.915966386554622
0.7 0.915966386554622
0.71 0.915966386554622
0.72 0.915966386554622
0.73 0.915966386554622
0.74 0.915966386554622
0.75 0.915966386554622
0.76 0.915966386554622
0.77 0.915966386554622
0.78 0.915966386554622
0.79 0.915966386554622
0.8 0.915966386554622
0.81 0.915966386554622
0.82 0.915966386554622
0.83 0.915966386554622
0.84 0.915966386554622
0.85 0.915966386554622
0.86 0.915966386554622
0.87 0.915966386554622
0.88 0.915966386554622
0.89 0.915966386554622
0.9 0.915966386554622
0.91 0.915966386554622
0.92 0.915966386554622
0.93 0.915966386554622
0.94 0.915966386554622
0.95 0.915966386554622
0.96 0.915966386554622
0.97 0.915966386554622
0.98 0.915966386554622
0.99 0.915966386554622
};
\addplot [line width=2.0pt, color1, opacity=0.8, dashed]
table {%
0 1
0.01 0.974380871050384
0.02 0.953031596925704
0.03 0.92997438087105
0.04 0.893253629376601
0.05 0.867634500426986
0.06 0.822374039282664
0.07 0.800170794192997
0.08 0.78736122971819
0.09 0.755764304013664
0.1 0.728437233134073
0.11 0.700256191289496
0.12 0.671221178479932
0.13 0.649017933390265
0.14 0.649017933390265
0.15 0.613151152860803
0.16 0.613151152860803
0.17 0.552519214346712
0.18 0.552519214346712
0.19 0.552519214346712
0.2 0.552519214346712
0.21 0.494449188727583
0.22 0.494449188727583
0.23 0.494449188727583
0.24 0.494449188727583
0.25 0.494449188727583
0.26 0.394534585824082
0.27 0.394534585824082
0.28 0.394534585824082
0.29 0.394534585824082
0.3 0.394534585824082
0.31 0.394534585824082
0.32 0.394534585824082
0.33 0.394534585824082
0.34 0.257899231426132
0.35 0.257899231426132
0.36 0.257899231426132
0.37 0.257899231426132
0.38 0.257899231426132
0.39 0.257899231426132
0.4 0.257899231426132
0.41 0.257045260461144
0.42 0.257045260461144
0.43 0.257045260461144
0.44 0.257045260461144
0.45 0.257045260461144
0.46 0.257045260461144
0.47 0.257045260461144
0.48 0.257045260461144
0.49 0.257045260461144
0.5 0.257045260461144
0.51 0.0930828351836038
0.52 0.0930828351836038
0.53 0.0930828351836038
0.54 0.0930828351836038
0.55 0.0930828351836038
0.56 0.0930828351836038
0.57 0.0930828351836038
0.58 0.0930828351836038
0.59 0.0930828351836038
0.6 0.0930828351836038
0.61 0.0930828351836038
0.62 0.0930828351836038
0.63 0.0930828351836038
0.64 0.0930828351836038
0.65 0.0930828351836038
0.66 0.0930828351836038
0.67 0.0930828351836038
0.68 0.0930828351836038
0.69 0.0930828351836038
0.7 0.0930828351836038
0.71 0.0930828351836038
0.72 0.0930828351836038
0.73 0.0930828351836038
0.74 0.0930828351836038
0.75 0.0930828351836038
0.76 0.0930828351836038
0.77 0.0930828351836038
0.78 0.0930828351836038
0.79 0.0930828351836038
0.8 0.0930828351836038
0.81 0.0930828351836038
0.82 0.0930828351836038
0.83 0.0930828351836038
0.84 0.0930828351836038
0.85 0.0930828351836038
0.86 0.0930828351836038
0.87 0.0930828351836038
0.88 0.0930828351836038
0.89 0.0930828351836038
0.9 0.0930828351836038
0.91 0.0930828351836038
0.92 0.0930828351836038
0.93 0.0930828351836038
0.94 0.0930828351836038
0.95 0.0930828351836038
0.96 0.0930828351836038
0.97 0.0930828351836038
0.98 0.0930828351836038
0.99 0.0930828351836038
};
\end{axis}

\node at (0.5,6.25) {$E_2:$};

\end{tikzpicture}}
	\end{subfigure}
	\caption{Panel (a):  $D_{\delta_r}/D$ (dashed blue) and $D_{\delta_r}/E_{\delta_r}$ (solid yellow). The monotonic decrease of the ratio among disputed entities and entities with response distance $\geq \delta_r$ indicates that disputed entities do not generate much debates and critics from their audience. Panel (b): $D_{\rho_e}/D$ (dashed blue) and $D_{\rho_e}/E_{\rho_e}$ (solid green). The monotonic increase of the ratio among disputed entities and entities with user share $\geq \rho_e$ indicates that most viral posts can reach up to $\sim 90\%$ of the basin of possible users.}
	\label{fig:response}
\end{figure}
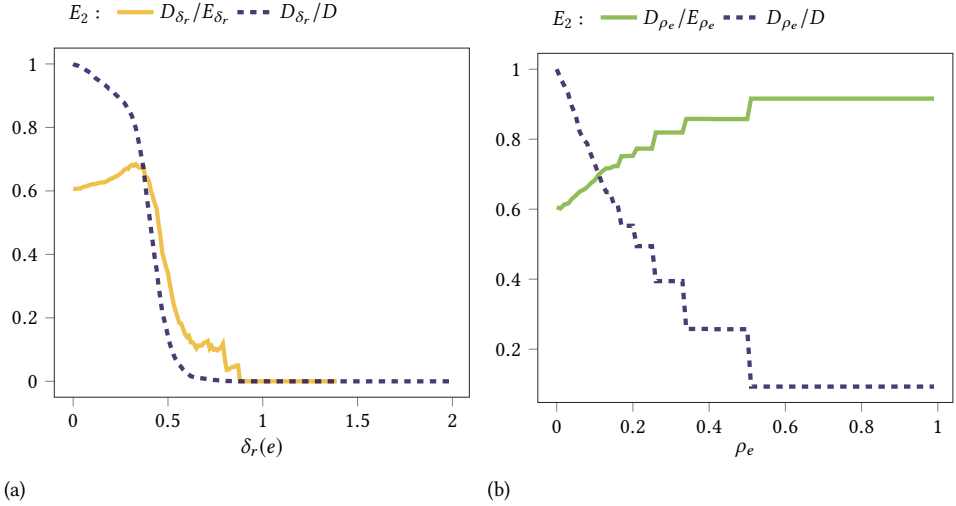

In Figure~\ref{fig:response} (a) we show the fraction $D_{\delta_r}/D$ of disputed entities with response distance $d_r$ greater than or equal to $\delta_r$ and the ratio $D_{\delta_r}/E_{\delta_r}$ among disputed entities with $d_r\ge\delta_r$ and entities with the same $d_r\ge\delta_r$. By definition, $D_{\delta_r}/D$ is a decreasing function of $\delta_r$ but this time it is not present an evident plateau indicating a clearcut division into two distinct subsets. Also, unlike Figure~\ref{fig:pres_dist}, the quantity $D_{\delta_r}/E_{\delta_r}$ is monotonically decreasing. Such a behavior indicates that the higher the presentation distance, the lower the probability that such post is disputed, i.e. there is a higher probability to have disputed entities -- and hence a higher probability of dealing with fake news -- when the users' response is consonant with the presentation given by news sources.  The quantity $D_{\delta_r}/E_{\delta_r}$ starts from an initial value of $\sim 0.6$  that is followed by a sudden downfall.  

In Figure~\ref{fig:response} (b) we show the fraction $D_{\rho_e}/D$ of disputed entities engaging a fraction of users $\geq \rho_e$ and the ratio $D_{\rho_e}/E_{\rho_e}$ among disputed entities and entities engaging a fraction of users $\geq \rho_e$. We observe that again $D_{\rho_e}/D$ does not show a clearcut plateau. The same lack of a plateau happens for $D_{\rho_e}/E_{\rho_e}$, that this time is an increasing quantity. Such result indicates that the higher the share of engaged users, the higher the probability that a post is debated. In particular, we can read from the curve ($\rho_e \to 1$) that $\sim 90\%$ of the most viral posts is disputed.

Notice that since we observe a monotonic behavior, the inflection point is just a proxy of the separation point among two subsets fo entities with high and low values of the analyzed threshold parameter. By fitting our data to polynomial functions and computing all inflection points, we get the thresholds $\delta_r(e) = 0.27$ and $\rho_e = 0.42$.

\subsection{Classification}\label{sec:early_class}
Now that we have defined all the features, we are ready for the last step of our framework i.e., the classification task. Our aim is to identify entities that are potential targets for fake news. In other words, we want to assess if an entity, and hence a topic, found on a post has a high probability to be soon found on fake news as well. Table~\ref{tab:features_early} provides a list of all the features employed in our classifiers. Notice that the main difference consists in the fact that sample $E_2$ also benefits from features involving the sentiment score of users comments (features 6--11).

\begin{table}[tb]
	\centering
	\caption{\textbf{Early Warning.} Features.}
	\label{tab:features_early}
	\begin{tabular}{llll}
		\toprule
		ID & Feature name & N. of features & Sample\\
		\midrule
		1 & \textit{Occurrences} & $1$ &   $E_1, E_2$ \\
		2 & \textit{Min/Max/Mean/Std sentiment score on posts} & $4$ &  $E_1, E_2$ \\
		3 & \textit{Presentation distance}&  $1$ &  $E_1, E_2$  \\
		4 & \textit{Number of negative posts} &  $1$ &  $E_1, E_2$  \\
		5 & \textit{Controversy} &  $1$ &  $E_1, E_2$  \\
		6 & \textit{Min/Max/Mean/Std sentiment score on comments} & $4$ &   $E_2$ \\
		7 & \textit{Min/Max/Mean/Std response distance} &  $4$ &  $E_2$ \\
		8 & \textit{Comments count} &  $1$ & $E_2$\\
		9 & \textit{Number of negative comments} & $1$ &   $E_2$  \\
		10 &\textit{Perception} &   $1$ & $E_2$  \\
		11 &\textit{Captivation} &  $1$ &  $E_2$  \\
		\bottomrule
	\end{tabular}
\end{table}

As illustrated in Section~\ref{classification}, we compare the results of different classifiers: Linear Regression (LIN), Logistic Regression (LOG), Support Vector Machine (SVM) with linear kernel, K-Nearest Neighbors (KNN) with $K=5$, Neural Network Models (NN) through the Multi-layer Perceptron L-BFGS algorithm, and Decision Trees (DT) with Gini Index. Given the asymmetry of our dataset in favor of official news sources, we re-sample the data at each step in order to get two balanced groups. For the sake of simplicity, in Table~\ref{tab:classification_early} we report the classification results only for the four best performing algorithms. We may notice an appreciable high accuracy --especially for the case of data sample $E_1$-- and observe that all algorithms are able to accurately recognize undisputed topics, however their ability decreases in the case of disputed ones. Moreover, we notice a significantly low FP Rate for the case of disputed entities achieved by both LOG and NN, meaning that even though they are more difficult to detect, there is also a smaller probability to falsely label a disputed entity as not. 

\begin{table}[tb]
	\centering
	\caption{\textbf{Early Warning: Classification Results.}  We report the performances for the four best performing algorithms (LOG, SVM, KNN, NN). The first reported value refers to $E_2$, while that for $E_1$ is in parenthesis. Values in bold denote the two best algorithms. W. Avg. denotes the weighted average across the two classes.}
	\label{tab:classification_early} 
	\scalebox{0.8}{	\begin{tabular}{lllllllll}
			\toprule
			& AUC &Accuracy&M. Abs. Err.&	 &Precision& Recall& FP Rate&F1-score\\
			\midrule
			& & & &	{\textit{Undisputed}} &  $0.74 (0.77)$ & $0.95 (0.92)$ &$0.50 (0.42) $ & $0.83 (0.84)$ \\
			{\textit{LOG}} & \textbf{0.73 (0.76)} & $\textbf{0.77 (0.79)}$& $\textbf{0.23 (0.21)}$&	{\textit{Disputed}}   & $0.87 (0.84)$ & $0.50 (0.59)$ &$0.05 (0.08)$& $0.63 (0.69)$\\
			& & & &	{\textit{W. Avg.}}& $0.79 (0.80)$ & $0.77 (0.79)$ & $0.28 (0.25)$& $0.75 (0.78)$\\
			\midrule
			& & & &	{\textit{Undisputed}}   & $0.75 (0.81)$ & $0.77 (0.89)$ &$0.38 (0.32) $& $0.76 (0.85)$\\ 
			{\textit{SVM}} & 0.68 (0.74) & $0.71 (0.80)$& $0.29 (0.20)$ &	{\textit{Disputed}}  &$0.64 (0.80)$ & $0.62 (0.67)$ & $0.32 (0.11)$& $0.63 (0.73)$\\
			& & & &	{\textit{W. Avg.}} & $0.71 (0.80)$ & $0.71 (0.80)$ &$0.35 (0.21)$ & $0.71 (0.80)$\\
	       \midrule
			& & & &	{\textit{Undisputed}}   & $0.71 (0.80)$ & $0.77 (0.84)$ &$0.58 (0.31) $& $0.74 (0.82)$\\ 
			{\textit{KNN}} & 0.60 (0.75) & $0.67 (0.78)$& $0.33 (0.22)$ &	{\textit{Disputed}}   &$0.60 (0.74)$ & $0.53 (0.69)$ & $0.23 (0.16)$& $0.56 (0.71)$\\
			& & & &	{\textit{W. Avg.}} & $0.66 (0.78)$ & $0.67 (0.78)$  &$0.35 (0.23)$& $0.67 (0.78)$\\
			\midrule
			& & & &	{\textit{Undisputed}}    &$0.71 (0.79)$ & $0.92 (0.92)$&$0.57 (0.37) $ & $0.80 (0.85)$\\ 
			{\textit{NN}}  & \textbf{0.68 (0.77)} & $\textbf{0.72 (0.80)}$& $\textbf{0.28 (0.20)}$&	{\textit{Disputed}}  & $0.77 (0.84)$ & $0.42 (0.63)$ &$0.08 (0.08)$ & $0.55 (0.72)$\\
			& & & &	{\textit{W. Avg.}}    & $0.73 (0.81)$ & $0.72 (0.80)$ &$0.32 (0.22)$& $0.70 (0.80)$\\
			\bottomrule
	\end{tabular}}
\end{table}	

Once detected the two best algorithms --i.e., LOG and NN-- we use them to classify entities again. Specifically, we take the whole samples --$E_1$ and $E_2$-- and we make predictions about the potentiality of each entity to become object of fake news by using either LOG or NN. We then keep the two predicted values for each entity. Looking at the AUC score, we may determine the best performing features for our classifier. We use the forward stepwise features selection where, starting from an empty set of features, we iteratively add the best performing one among the unselected, when tested together with the best ones selected so far. Table~\ref{tab:bestfeatures_early} reports the best seven features, for $E_1$ and $E_2$, for both algorithms LOG and NN. Note that the newly introduced measures (in bold) --i.e., presentation distance, response distance, controversy, perception, and captivation-- are always among the best performing features. Moreover, we observe that the presentation distance is the best performing feature in all cases, with the exception of the logistic regression on $E_2$, where instead we find the response distance to be among the first three.

\begin{table}[tb]
	\centering
	\caption{\textbf{Early Warning.} Best performing features.}
	\label{tab:bestfeatures_early} 
	\scalebox{0.8}{	\begin{tabular}{lllll}
			\toprule
			&LOG& 	& NN &\\
			\midrule
			&$E_1$&$E_2$&$E_1$	& $E_2$ \\
			\midrule
			$1$ & \textbf{Presentation distance}& Occurrences&  \textbf{Presentation distance}& \textbf{Presentation distance} \\ 
			$2$ &\textbf{Captivation}& \textbf{Std response distance}& \textbf{Captivation} & Occurrences\\ 
			$3$ &  Mean post sent. score& \textbf{Min response distance} &Min post sent. score&\textbf{Captivation}\\ 
			$4$ &  \textbf{Controversy}& Std comm. sent. score&Max post sent. score& Mean comm. sent. score\\ 
			$5$ &  Min post sent. score& Min comm. sent. score&Std post sent. score&\textbf{Mean response distance}\\ 
			$6$ & Max post sent. score& \textbf{Captivation}  &Number negative posts&\textbf{Perception}\\ 
			$7$ & Number negative posts & Mean comm. sent. score&\textbf{Controversy}&\textbf{Max response distance}\\ 
			\bottomrule
	\end{tabular}}
\end{table}	

\subsection{Insights}
We observed that both presentation and response distance play a primary role in the classification of disputed topics and are among the best features of our classifier. Thus, it is worth studying in deep such measures to better understand users' behavior on the platform.

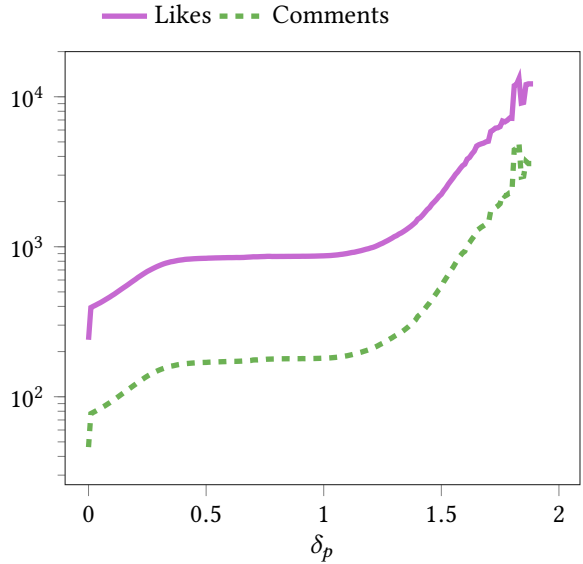
\begin{figure}[tb]
	\centering
\begin{tikzpicture}


\definecolor{color0}{RGB}{184,68,198}
\definecolor{color1}{RGB}{72,158,42}

\begin{axis}[
xlabel={$\delta_p$},
xmin=-0.0995, xmax=2.0895,
ymin=25.9152901876822, ymax=20000,
ymode=log,
tick align=outside,
tick pos=left,
x grid style={white!80.0!black},
y grid style={white!80.0!black},
axis line style={darkgray!60.0!black},
legend style={at={(0.05,1.03)}, anchor=south west, draw=none},
legend columns=2,
legend entries={{Likes},{Comments}}
]
\addplot [line width=2.0pt, color0, opacity=0.8]
table {%
	0 239.652225420509
	0.01 392.98451846391
	0.02 400.581956689207
	0.03 407.730944523024
	0.04 415.686055022941
	0.05 423.396488221136
	0.06 432.381210107872
	0.07 441.924304692727
	0.08 451.646286498138
	0.09 461.533669593875
	0.1 472.746096976706
	0.11 483.482841214999
	0.12 495.918735694823
	0.13 508.754880645523
	0.14 521.668381470887
	0.15 534.732253397529
	0.16 547.765563917123
	0.17 561.900993335848
	0.18 576.770070532103
	0.19 591.696261057808
	0.2 606.493155579989
	0.21 623.416789605084
	0.22 638.281212581408
	0.23 653.798424912829
	0.24 669.124992284427
	0.25 684.453026941363
	0.26 697.329801754956
	0.27 710.619129972063
	0.28 723.609861503386
	0.29 736.261961930612
	0.3 747.88528253698
	0.31 759.011627487091
	0.32 769.500036767409
	0.33 778.329863831375
	0.34 785.802618633164
	0.35 793.827781823757
	0.36 798.173608423008
	0.37 804.234712230216
	0.38 810.475767971923
	0.39 814.792073703439
	0.4 818.437906930298
	0.41 823.183138248663
	0.42 825.732940321408
	0.43 827.453753595592
	0.44 829.88310896002
	0.45 831.942042580961
	0.46 833.404867075665
	0.47 834.922423695438
	0.48 835.992316226322
	0.49 836.898637804025
	0.5 837.435762244099
	0.51 839.607576633892
	0.52 840.134500020686
	0.53 841.217308409666
	0.54 842.305205047319
	0.55 843.051618802211
	0.56 843.626128938278
	0.57 844.575174023592
	0.58 845.089239502463
	0.59 845.970392673441
	0.6 846.258437316808
	0.61 846.12010246504
	0.62 846.760460910888
	0.63 847.548680772149
	0.64 847.073554870582
	0.65 847.814935477047
	0.66 848.602112676056
	0.67 850.867659574468
	0.68 852.809804172533
	0.69 854.454996790071
	0.7 856.909408869122
	0.71 856.815356648199
	0.72 857.189503866539
	0.73 858.854521682284
	0.74 859.827925485394
	0.75 860.957861966149
	0.76 862.921065375303
	0.77 863.116712280392
	0.78 861.136936618161
	0.79 861.400017715576
	0.8 861.182426444523
	0.81 861.624138084434
	0.82 861.839403383793
	0.83 861.779223093371
	0.84 862.220850418969
	0.85 862.604191750279
	0.86 862.322868753628
	0.87 862.904138362531
	0.88 863.022574877068
	0.89 863.099740701001
	0.9 863.324076558447
	0.91 864.200752485891
	0.92 863.883038315035
	0.93 864.198116591928
	0.94 864.577582338688
	0.95 865.40013477089
	0.96 865.78436572867
	0.97 866.868073997389
	0.98 867.76196269262
	0.99 868.88111572486
	1 869.567838741752
	1.01 871.157555978606
	1.02 872.515528514348
	1.03 874.148544131028
	1.04 877.139132420091
	1.05 879.88780979431
	1.06 883.161576671879
	1.07 887.996945999722
	1.08 891.404639055455
	1.09 896.680806285661
	1.1 901.788775462201
	1.11 908.720387243736
	1.12 915.401244912617
	1.13 919.320255430313
	1.14 927.292310702407
	1.15 935.080601295555
	1.16 942.841513956159
	1.17 952.66471004959
	1.18 962.424104167735
	1.19 970.304472178887
	1.2 981.907618444655
	1.21 991.252745478036
	1.22 1005.05127362366
	1.23 1021.59051820415
	1.24 1039.91648577631
	1.25 1055.56209531587
	1.26 1076.41136023916
	1.27 1095.86662177878
	1.28 1117.82411302983
	1.29 1140.69809499292
	1.3 1167.3017133029
	1.31 1186.36487221316
	1.32 1215.49663205164
	1.33 1237.12806637807
	1.34 1265.5736700538
	1.35 1296.54396112911
	1.36 1331.32412080429
	1.37 1366.87168326777
	1.38 1409.12583936513
	1.39 1453.56951408515
	1.4 1528.34788526981
	1.41 1561.74328960646
	1.42 1621.34902958956
	1.43 1692.26311040938
	1.44 1767.96115453999
	1.45 1825.32479716024
	1.46 1920.57249931862
	1.47 1991.66647204788
	1.48 2070.27372034956
	1.49 2165.67953216374
	1.5 2236.87646544182
	1.51 2362.43349657199
	1.52 2473.54787345485
	1.53 2617.7196557971
	1.54 2745.30214948705
	1.55 2878.11645838873
	1.56 3033.96175963197
	1.57 3158.17118802619
	1.58 3319.05959156344
	1.59 3466.23154121864
	1.6 3554.66037735849
	1.61 3839.74257425743
	1.62 3939.76131117267
	1.63 4156.38762376238
	1.64 4338.39253333333
	1.65 4691.21515892421
	1.66 4791.0101010101
	1.67 4855.80147058824
	1.68 4910.45340751043
	1.69 5007.01436781609
	1.7 5052.58013052937
	1.71 5859.12909441233
	1.72 5997.94550958628
	1.73 6176.0158061117
	1.74 6211.38138479001
	1.75 6325.6305125149
	1.76 6890.55587808418
	1.77 6802.51446945338
	1.78 6955.72277227723
	1.79 7188.54700854701
	1.8 7180.35800344234
	1.81 11850.5289855072
	1.82 12113.8740740741
	1.83 13193.9661016949
	1.84 9079.90825688073
	1.85 9163.98148148148
	1.86 12050
	1.87 12187.5384615385
	1.88 12187.5384615385
	1.89 12187.5384615385
	1.9 0
	1.91 0
	1.92 0
	1.93 0
	1.94 0
	1.95 0
	1.96 0
	1.97 0
	1.98 0
	1.99 0
};
\addplot [line width=2.0pt, color1, opacity=0.8, dashed]
table {%
	0 46.1333979517907
	0.01 77.1598418336573
	0.02 78.6587349139776
	0.03 80.1610176890705
	0.04 81.8444646778426
	0.05 83.4372584056597
	0.06 85.2414430219103
	0.07 87.0579948389563
	0.08 89.1391338428376
	0.09 91.2030022305729
	0.1 93.5452461589295
	0.11 95.7770735739873
	0.12 98.3279564032698
	0.13 100.972677350667
	0.14 103.68455292055
	0.15 106.391480682115
	0.16 109.149017787334
	0.17 112.123173645165
	0.18 115.05354734986
	0.19 118.153860544672
	0.2 121.179274326553
	0.21 124.517107353609
	0.22 127.798487193715
	0.23 130.830377539978
	0.24 134.114283068946
	0.25 137.210110935024
	0.26 140.070165745856
	0.27 142.887255554077
	0.28 145.58750467894
	0.29 148.117340014615
	0.3 150.477244510659
	0.31 152.679846892717
	0.32 154.617802779616
	0.33 156.400783435926
	0.34 158.19015363657
	0.35 159.880141055617
	0.36 160.836378416041
	0.37 162.186072880826
	0.38 163.466619346189
	0.39 164.45413390517
	0.4 164.976712602357
	0.41 166.022967700414
	0.42 166.547282564807
	0.43 167.050196491512
	0.44 167.592670923659
	0.45 168.022432498572
	0.46 168.419345603272
	0.47 168.833565802429
	0.48 169.084028434072
	0.49 169.293386559118
	0.5 169.3936647856
	0.51 169.870817152772
	0.52 170.103264242274
	0.53 170.313300451776
	0.54 170.551344844762
	0.55 170.757242009892
	0.56 170.981187830358
	0.57 171.224542536785
	0.58 171.413473578763
	0.59 171.69890854347
	0.6 171.692446193786
	0.61 172.05207239743
	0.62 172.067832961857
	0.63 172.209727724859
	0.64 172.434615547017
	0.65 172.736788660884
	0.66 173.116494145596
	0.67 173.607361702128
	0.68 173.991936516063
	0.69 174.468992082174
	0.7 175.772010881299
	0.71 176.081890581717
	0.72 175.968807020593
	0.73 176.459751579865
	0.74 176.974593318174
	0.75 177.304262036306
	0.76 177.918688091569
	0.77 178.118786969189
	0.78 178.337033924543
	0.79 178.456087514948
	0.8 178.522952853598
	0.81 178.883357800614
	0.82 179.011175422974
	0.83 179.064994654312
	0.84 179.159163843822
	0.85 179.227469342252
	0.86 179.09949537802
	0.87 179.230380765105
	0.88 179.260661600358
	0.89 179.276689914163
	0.9 179.324702620517
	0.91 179.594374272149
	0.92 179.335200537755
	0.93 179.449820627803
	0.94 179.511487032218
	0.95 179.71832884097
	0.96 179.81718061674
	0.97 180.045505693838
	0.98 180.230332522303
	0.99 180.492327134862
	1 180.715312302269
	1.01 181.010923760312
	1.02 181.318788594261
	1.03 181.677934485896
	1.04 182.323515981735
	1.05 182.908607815292
	1.06 183.61972219667
	1.07 184.68627088057
	1.08 185.465857853391
	1.09 186.585632775232
	1.1 187.651644164275
	1.11 189.197845482156
	1.12 190.777783097917
	1.13 192.517681776402
	1.14 194.516581882215
	1.15 196.471393957375
	1.16 198.293204174365
	1.17 200.693249671086
	1.18 202.924642436049
	1.19 205.349193967759
	1.2 207.917115120819
	1.21 211.409399224806
	1.22 214.77063818132
	1.23 218.706723252178
	1.24 223.163543927796
	1.25 227.089424254798
	1.26 231.870612855007
	1.27 236.997184305564
	1.28 241.966656200942
	1.29 246.913888531343
	1.3 252.91751008798
	1.31 258.150761283306
	1.32 264.872088127982
	1.33 268.991558441558
	1.34 276.417364016736
	1.35 284.357936140676
	1.36 292.833373387807
	1.37 303.269021511677
	1.38 313.135344902764
	1.39 324.644908378157
	1.4 343.480700048614
	1.41 354.292230070636
	1.42 367.662424435253
	1.43 386.159242133754
	1.44 405.039927841251
	1.45 422.766734279919
	1.46 447.203870264377
	1.47 467.954751131222
	1.48 492.424313358302
	1.49 519.144193817878
	1.5 537.085914260717
	1.51 582.60430950049
	1.52 616.340037712131
	1.53 652.09669384058
	1.54 692.936003908158
	1.55 729.224408402021
	1.56 775.569867740081
	1.57 822.980979108201
	1.58 867.858051556746
	1.59 907.982078853047
	1.6 932.80244173141
	1.61 1024.09556607835
	1.62 1066.23407202216
	1.63 1124.19306930693
	1.64 1179.58773333333
	1.65 1278.81907090465
	1.66 1313.94507575758
	1.67 1351.48529411765
	1.68 1381.04033379694
	1.69 1416.07112068966
	1.7 1429.18999274837
	1.71 1731.54527938343
	1.72 1796.61049445005
	1.73 1822.95363540569
	1.74 1868.86379114642
	1.75 1937.34922526818
	1.76 2243.10449927431
	1.77 2176.32154340836
	1.78 2219.79372937294
	1.79 2284.79658119658
	1.8 2293.27710843374
	1.81 4427.73188405797
	1.82 4526.12592592593
	1.83 4950.40677966102
	1.84 2923.53211009174
	1.85 2950.60185185185
	1.86 3708.19718309859
	1.87 3586.18461538462
	1.88 3586.18461538462
	1.89 3586.18461538462
	1.9 0
	1.91 0
	1.92 0
	1.93 0
	1.94 0
	1.95 0
	1.96 0
	1.97 0
	1.98 0
	1.99 0
};
\end{axis}

\end{tikzpicture}
\caption{Mean number of likes (solid purple) and comments (dashed green) for entities whose presentation distance $d_p(e)\geq\delta_p$.}\label{fig:virality}
\end{figure}

Looking at Figure~~\ref{fig:virality}, we may observe a positive relationship between an entity's presentation distance and the attention received in terms of likes and comments. Indeed, entities that are presented in a different way by official and fake news sources get on average a higher number of likes and comments and, hence, a higher attention. 

\begin{figure}[tb]
	\centering
	\input{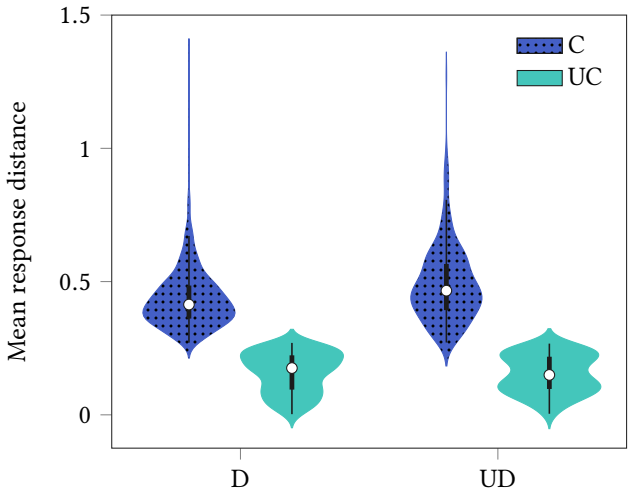}
	\caption{Estimated probability density function of the mean response distance for controversial (C) (dotted violet) and uncontroversial (UC) (aquamarine) entities by disputed (D) and undisputed (UD) entities.}\label{fig:controversy_comparison}
\end{figure}

Figure~\ref{fig:controversy_comparison} shows a series of violin plots representing the estimated probability density function of the mean response distance. The measure is computed for the two classes of entities --controversial (C) and uncontroversial (UC)-- and for both disputed (D) and undisputed (UD) entities. We may notice some significant differences: distributions of controversial entities show a main peak around $0.4$, which is larger in the case of disputed entities and followed by a smaller peak around $0.6$, whereas distributions of uncontroversial entities are centered around smaller values and present two peaks, which are of similar size in the case of disputed entities. This evidence suggests that for controversial entities users' response is usually divergent (mean response distance near to $0.4$) from the post presentation, while for uncontroversial ones there are mixed responses, that can be either similar to the presentation (mean response distance near to $0$) or slightly divergent (mean response distance near to $0.25$). Also, it may indicate that there is a higher probability of divergent response for disputed entities.

\begin{figure}[tb]
	\centering
    \input{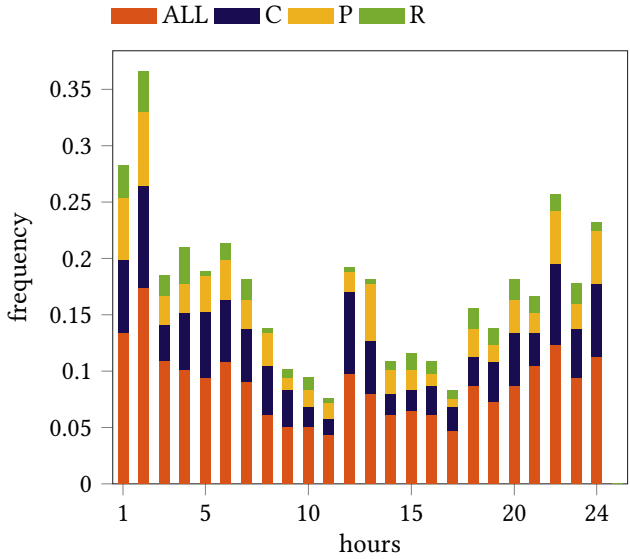}
	\caption{Frequency of the temporal distance between the first appearance of an entity and the consequent first appearance of an hoax containing that entity, for the following classes of entities: all (ALL), controversial (C), arousing controversial response (R), and captivating (P).}\label{fig:temp_fake}
\end{figure}

Finally, Figure~\ref{fig:temp_fake} shows the Frequency of the temporal distance between the first appearance of an entity in official information and the consequent first appearance in a fake news. Data refer to the following categories of entities: all (ALL), controversial (C), arousing controversial response (R), captivating (P). We may observe that the emergence of an entity on fake news is confined to about $24$ hours after its first appearance into official news. Moreover, it is worth noting that out of about $2K$ unique entities appearing on official newspapers, about $50\%$ is also the subject of either a hoax or fake news, and the same percentage of entities shows up on fake news only after a first apparison on official news. 

\section{Can we detect Fake News?}\label{fakedetection}
As we have seen, we are now able to issue a warning about topics that require special caution. This could be a crucial element of a broader system aiming at monitoring information flow constantly and in real time. As a further step, we want to exploit the output of our framework --i.e., whether a topic would appear in fake news or not-- to build a new feature set for a classification task with the aim of distinguishing fake news from reliable information. To this end, we show a possible application to Facebook data. As always, the approach is easily reproducible and suitable to other social media platforms.  

\subsection{Instantiation to Italian News on Facebook}
Let us consider two samples, $P_1$ and $P_2$, that are built from $E_1$ and $E_2$ (see Section \ref{sec:entities}) by taking all the news --i.e. the posts-- containing such entities. More specifically, let $\mathcal{P}$ be the set of all posts in our dataset. We define: $P_i =\{p\in \mathcal{P} : \exists e\in E_p | e\in E_i\}$ for $i \in \{1,2\}$
where $E_p$ represents the set of all entities contained in post $p$. As before, the main difference between the samples is the availability of sentiment scores for comments in the second one. 

\begin{figure}[tb]	
		\begin{subfigure}[]	
		\centering
		\resizebox{0.45\linewidth}{!}{
\begin{tikzpicture}

\definecolor{color0}{rgb}{0.85,0.325,0.098}
\definecolor{color1}{rgb}{0.929,0.694,0.125}
\definecolor{color2}{rgb}{0.101960784313725,0.0470588235294118,0.317647058823529}
\definecolor{color3}{rgb}{0.466,0.674,0.188}

\begin{axis}[
ylabel={$P_1$},
xmin=-0.2625, xmax=3.8625,
ymin=0.7, ymax=0.95,
xtick={0.3,1.3,2.3,3.3},
xticklabels={LIN,LOG,KNN,DT},
tick align=outside,
xticklabel style = {rotate=90},
tick pos=left,
x grid style={white!80.0!black},
y grid style={white!80.0!black},
axis line style={darkgray!60.0!black},
legend entries={{ST},{ST+S},{ST+UB},{ST+SB},{ST+P}},
legend cell align={left},
legend style={at={(0.05,1.05)}, anchor=south west, draw=none},
legend columns=2
]
\addlegendimage{area legend,draw=black,fill=color0,postaction={pattern=vertical lines}};
\draw[draw=black,fill=color0,postaction={pattern=vertical lines}] (axis cs:-0.075,0) rectangle (axis cs:0.075,0.724705524614301);
\draw[draw=black,fill=color0,postaction={pattern=vertical lines}] (axis cs:0.925,0) rectangle (axis cs:1.075,0.778407272261699);
\draw[draw=black,fill=color0,postaction={pattern=vertical lines}] (axis cs:1.925,0) rectangle (axis cs:2.075,0.852346200627361);
\draw[draw=black,fill=color0,postaction={pattern=vertical lines}] (axis cs:2.925,0) rectangle (axis cs:3.075,0.824499071762371);
\addlegendimage{area legend,draw=black,fill=blue,postaction={pattern=grid}};
\draw[draw=black,fill=blue,postaction={pattern=grid}] (axis cs:0.075,0) rectangle (axis cs:0.225,0.809735292234812);
\draw[draw=black,fill=blue,postaction={pattern=grid}] (axis cs:1.075,0) rectangle (axis cs:1.225,0.85158600601754);
\draw[draw=black,fill=blue,postaction={pattern=grid}] (axis cs:2.075,0) rectangle (axis cs:2.225,0.862700851417963);
\draw[draw=black,fill=blue,postaction={pattern=grid}] (axis cs:3.075,0) rectangle (axis cs:3.225,0.841135330644645);
\addlegendimage{area legend,draw=black,fill=color1,postaction={pattern=dots}};
\draw[draw=black,fill=color1,postaction={pattern=dots}] (axis cs:0.225,0) rectangle (axis cs:0.375,0.797460149798348);
\draw[draw=black,fill=color1,postaction={pattern=dots}] (axis cs:1.225,0) rectangle (axis cs:1.375,0.836966263363421);
\draw[draw=black,fill=color1,postaction={pattern=dots}] (axis cs:2.225,0) rectangle (axis cs:2.375,0.845864541322578);
\draw[draw=black,fill=color1,postaction={pattern=dots}] (axis cs:3.225,0) rectangle (axis cs:3.375,0.820746111004417);
\addlegendimage{area legend,draw=black,fill=color2,draw opacity=0};
\draw[draw=black,fill=color2] (axis cs:0.375,0) rectangle (axis cs:0.525,0.732427501440369);
\draw[draw=black,fill=color2] (axis cs:1.375,0) rectangle (axis cs:1.525,0.794827475833814);
\draw[draw=black,fill=color2] (axis cs:2.375,0) rectangle (axis cs:2.525,0.854346712758466);
\draw[draw=black,fill=color2] (axis cs:3.375,0) rectangle (axis cs:3.525,0.806590487164714);
\addlegendimage{area legend,draw=black,fill=color3,postaction={pattern=horizontal lines}};
\draw[draw=black,fill=color3,postaction={pattern=horizontal lines}] (axis cs:0.525,0) rectangle (axis cs:0.675,0.733699827155752);
\draw[draw=black,fill=color3,postaction={pattern=horizontal lines}] (axis cs:1.525,0) rectangle (axis cs:1.675,0.794891492222009);
\draw[draw=black,fill=color3,postaction={pattern=horizontal lines}] (axis cs:2.525,0) rectangle (axis cs:2.675,0.858219704244287);
\draw[draw=black,fill=color3,postaction={pattern=horizontal lines}] (axis cs:3.525,0) rectangle (axis cs:3.675,0.828276038665898);
\end{axis}

\end{tikzpicture}}
	\end{subfigure}
	\begin{subfigure}[]	
		\centering
		\resizebox{0.45\linewidth}{!}{
\begin{tikzpicture}

\definecolor{color0}{rgb}{0.85,0.325,0.098}
\definecolor{color1}{rgb}{0.929,0.694,0.125}
\definecolor{color2}{rgb}{0.101960784313725,0.0470588235294118,0.317647058823529}
\definecolor{color3}{rgb}{0.466,0.674,0.188}

\begin{axis}[
ylabel={$P_2$},
xmin=-0.2625, xmax=3.8625,
ymin=0.7, ymax=0.95,
xtick={0.3,1.3,2.3,3.3},
xticklabels={LIN,LOG,KNN,DT},
tick align=outside,
xticklabel style = {rotate=90},
tick pos=left,
x grid style={white!80.0!black},
y grid style={white!80.0!black},
axis line style={darkgray!60.0!black},
legend entries={{ST},{ST+S},{ST+UB},{ST+SB},{ST+P}},
legend cell align={left},
legend style={at={(0.05,1.05)}, anchor=south west, draw=none},
legend columns=2
]
\addlegendimage{area legend,draw=black,fill=color0,postaction={pattern=vertical lines}};
\draw[draw=black,fill=color0,postaction={pattern=vertical lines}] (axis cs:-0.075,0) rectangle (axis cs:0.075,0.712713886978682);
\draw[draw=black,fill=color0,postaction={pattern=vertical lines}] (axis cs:0.925,0) rectangle (axis cs:1.075,0.793758343109098);
\draw[draw=black,fill=color0,postaction={pattern=vertical lines}] (axis cs:1.925,0) rectangle (axis cs:2.075,0.859662724898759);
\draw[draw=black,fill=color0,postaction={pattern=vertical lines}] (axis cs:2.925,0) rectangle (axis cs:3.075,0.80810736710579);
\addlegendimage{area legend,draw=black,fill=blue,postaction={pattern=grid}};
\draw[draw=black,fill=blue,postaction={pattern=grid}] (axis cs:0.075,0) rectangle (axis cs:0.225,0.841753974353788);
\draw[draw=black,fill=blue,postaction={pattern=grid}] (axis cs:1.075,0) rectangle (axis cs:1.225,0.884648229800125);
\draw[draw=black,fill=blue,postaction={pattern=grid}] (axis cs:2.075,0) rectangle (axis cs:2.225,0.89934109102023);
\draw[draw=black,fill=blue,postaction={pattern=grid}] (axis cs:3.075,0) rectangle (axis cs:3.225,0.86480904673939);
\addlegendimage{area legend,draw=black,fill=color1,postaction={pattern=dots}};
\draw[draw=black,fill=color1,postaction={pattern=dots}] (axis cs:0.225,0) rectangle (axis cs:0.375,0.834061459218209);
\draw[draw=black,fill=color1,postaction={pattern=dots}] (axis cs:1.225,0) rectangle (axis cs:1.375,0.873488127502933);
\draw[draw=black,fill=color1,postaction={pattern=dots}] (axis cs:2.225,0) rectangle (axis cs:2.375,0.846489039907591);
\draw[draw=black,fill=color1,postaction={pattern=dots}] (axis cs:3.225,0) rectangle (axis cs:3.375,0.854386059535528);
\addlegendimage{area legend,draw=black,fill=color2,draw opacity=0};
\draw[draw=black,fill=color2] (axis cs:0.375,0) rectangle (axis cs:0.525,0.799008040846963);
\draw[draw=black,fill=color2] (axis cs:1.375,0) rectangle (axis cs:1.525,0.86299996853777);
\draw[draw=black,fill=color2] (axis cs:2.375,0) rectangle (axis cs:2.525,0.868267644692547);
\draw[draw=black,fill=color2] (axis cs:3.375,0) rectangle (axis cs:3.525,0.862114531504928);
\addlegendimage{area legend,draw=black,fill=color3,postaction={pattern=horizontal lines}};
\draw[draw=black,fill=color3,postaction={pattern=horizontal lines}] (axis cs:0.525,0) rectangle (axis cs:0.675,0.804424038941251);
\draw[draw=black,fill=color3,postaction={pattern=horizontal lines}] (axis cs:1.525,0) rectangle (axis cs:1.675,0.848372279078966);
\draw[draw=black,fill=color3,postaction={pattern=horizontal lines}] (axis cs:2.525,0) rectangle (axis cs:2.675,0.886540458179955);
\draw[draw=black,fill=color3,postaction={pattern=horizontal lines}] (axis cs:3.525,0) rectangle (axis cs:3.675,0.861294266233387);
\end{axis}

\end{tikzpicture}}
	\end{subfigure}

	\begin{subfigure}[]	
		\centering
		\resizebox{0.45\linewidth}{!}{
\begin{tikzpicture}

\definecolor{color0}{rgb}{0.85,0.325,0.098}
\definecolor{color1}{rgb}{0.929,0.694,0.125}
\definecolor{color2}{rgb}{0.101960784313725,0.0470588235294118,0.317647058823529}
\definecolor{color3}{rgb}{0.466,0.674,0.188}

\begin{axis}[
ylabel={$P_1$},
xmin=-0.2625, xmax=3.8625,
ymin=0.7, ymax=0.95,
xtick={0.3,1.3,2.3,3.3},
xticklabels={LIN,LOG,KNN,DT},
tick align=outside,
xticklabel style = {rotate=90},
tick pos=left,
x grid style={white!80.0!black},
y grid style={white!80.0!black},
axis line style={darkgray!60.0!black},
legend entries={{ST},{ST+S},{ST+S+UB},{ST+S+UB+SB},{ST+S+UB+SB+P}},
legend cell align={left},
legend style={at={(0.05,1.05)}, anchor=south west, draw=none},
legend columns=2
]
\addlegendimage{area legend,draw=black,fill=color0,postaction={pattern=vertical lines}};
\draw[draw=black,fill=color0,postaction={pattern=vertical lines}] (axis cs:-0.075,0) rectangle (axis cs:0.075,0.712541287158443);
\draw[draw=black,fill=color0,postaction={pattern=vertical lines}] (axis cs:0.925,0) rectangle (axis cs:1.075,0.797727955159644);
\draw[draw=black,fill=color0,postaction={pattern=vertical lines}] (axis cs:1.925,0) rectangle (axis cs:2.075,0.855139625663097);
\draw[draw=black,fill=color0,postaction={pattern=vertical lines}] (axis cs:2.925,0) rectangle (axis cs:3.075,0.800120108097288);
\addlegendimage{area legend,draw=black,fill=blue,postaction={pattern=grid}};
\draw[draw=black,fill=blue,postaction={pattern=grid}] (axis cs:0.075,0) rectangle (axis cs:0.225,0.817715944349915);
\draw[draw=black,fill=blue,postaction={pattern=grid}] (axis cs:1.075,0) rectangle (axis cs:1.225,0.871524371934741);
\draw[draw=black,fill=blue,postaction={pattern=grid}] (axis cs:2.075,0) rectangle (axis cs:2.225,0.874857371634471);
\draw[draw=black,fill=blue,postaction={pattern=grid}] (axis cs:3.075,0) rectangle (axis cs:3.225,0.860234210789711);
\addlegendimage{area legend,draw=black,fill=color1,postaction={pattern=dots}};
\draw[draw=black,fill=color1,postaction={pattern=dots}] (axis cs:0.225,0) rectangle (axis cs:0.375,0.841247122410169);
\draw[draw=black,fill=color1,postaction={pattern=dots}] (axis cs:1.225,0) rectangle (axis cs:1.375,0.877159443499149);
\draw[draw=black,fill=color1,postaction={pattern=dots}] (axis cs:2.225,0) rectangle (axis cs:2.375,0.85890301271144);
\draw[draw=black,fill=color1,postaction={pattern=dots}] (axis cs:3.225,0) rectangle (axis cs:3.375,0.844339905915324);
\addlegendimage{area legend,fill=color2,draw opacity=0};
\draw[fill=color2,draw opacity=0] (axis cs:0.375,0) rectangle (axis cs:0.525,0.866970273245921);
\draw[fill=color2,draw opacity=0] (axis cs:1.375,0) rectangle (axis cs:1.525,0.885897307576819);
\draw[fill=color2,draw opacity=0] (axis cs:2.375,0) rectangle (axis cs:2.525,0.85890301271144);
\draw[fill=color2,draw opacity=0] (axis cs:3.375,0) rectangle (axis cs:3.525,0.844099689720749);
\addlegendimage{area legend,draw=black,fill=color3,postaction={pattern=horizontal lines}};
\draw[draw=black,fill=color3,postaction={pattern=horizontal lines}] (axis cs:0.525,0) rectangle (axis cs:0.675,0.867390651586428);
\draw[draw=black,fill=color3,postaction={pattern=horizontal lines}] (axis cs:1.525,0) rectangle (axis cs:1.675,0.889960964868381);
\draw[draw=black,fill=color3,postaction={pattern=horizontal lines}] (axis cs:2.525,0) rectangle (axis cs:2.675,0.861024922430187);
\draw[draw=black,fill=color3,postaction={pattern=horizontal lines}] (axis cs:3.525,0) rectangle (axis cs:3.675,0.832639375437894);
\end{axis}

\end{tikzpicture}}
   \end{subfigure}
	\begin{subfigure}[]	
		\centering
		\resizebox{0.45\linewidth}{!}{
\begin{tikzpicture}

\definecolor{color0}{rgb}{0.85,0.325,0.098}
\definecolor{color1}{rgb}{0.929,0.694,0.125}
\definecolor{color2}{rgb}{0.101960784313725,0.0470588235294118,0.317647058823529}
\definecolor{color3}{rgb}{0.466,0.674,0.188}

\begin{axis}[
ylabel={$P_2$},
xmin=-0.2625, xmax=3.8625,
ymin=0.7, ymax=0.95,
xtick={0.3,1.3,2.3,3.3},
xticklabels={LIN,LOG,KNN,DT},
tick align=outside,
xticklabel style = {rotate=90},
tick pos=left,
x grid style={white!80.0!black},
y grid style={white!80.0!black},
axis line style={darkgray!60.0!black},
legend entries={{ST},{ST+S},{ST+S+UB},{ST+S+UB+SB},{ST+S+UB+SB+P}},
legend cell align={left},
legend style={at={(0.05,1.05)}, anchor=south west, draw=none},
legend columns=2
]
\addlegendimage{area legend,draw=black,fill=color0,postaction={pattern=vertical lines}};
\draw[draw=black,fill=color0,postaction={pattern=vertical lines}] (axis cs:-0.075,0) rectangle (axis cs:0.075,0.760444444444444);
\draw[draw=black,fill=color0,postaction={pattern=vertical lines}] (axis cs:0.925,0) rectangle (axis cs:1.075,0.809333333333333);
\draw[draw=black,fill=color0,postaction={pattern=vertical lines}] (axis cs:1.925,0) rectangle (axis cs:2.075,0.838777777777778);
\draw[draw=black,fill=color0,postaction={pattern=vertical lines}] (axis cs:2.925,0) rectangle (axis cs:3.075,0.811777777777778);
\addlegendimage{area legend,draw=black,fill=blue,postaction={pattern=grid}};
\draw[draw=black,fill=blue,postaction={pattern=grid}] (axis cs:0.075,0) rectangle (axis cs:0.225,0.870888888888889);
\draw[draw=black,fill=blue,postaction={pattern=grid}] (axis cs:1.075,0) rectangle (axis cs:1.225,0.899111111111111);
\draw[draw=black,fill=blue,postaction={pattern=grid}] (axis cs:2.075,0) rectangle (axis cs:2.225,0.887444444444444);
\draw[draw=black,fill=blue,postaction={pattern=grid}] (axis cs:3.075,0) rectangle (axis cs:3.225,0.867777777777778);
\addlegendimage{area legend,draw=black,fill=color1,postaction={pattern=dots}};
\draw[draw=black,fill=color1,postaction={pattern=dots}] (axis cs:0.225,0) rectangle (axis cs:0.375,0.885111111111111);
\draw[draw=black,fill=color1,postaction={pattern=dots}] (axis cs:1.225,0) rectangle (axis cs:1.375,0.914111111111111);
\draw[draw=black,fill=color1,postaction={pattern=dots}] (axis cs:2.225,0) rectangle (axis cs:2.375,0.882222222222222);
\draw[draw=black,fill=color1,postaction={pattern=dots}] (axis cs:3.225,0) rectangle (axis cs:3.375,0.896333333333333);
\addlegendimage{area legend,fill=color2,draw opacity=0};
\draw[fill=color2,draw opacity=0] (axis cs:0.375,0) rectangle (axis cs:0.525,0.900444444444444);
\draw[fill=color2,draw opacity=0] (axis cs:1.375,0) rectangle (axis cs:1.525,0.937333333333334);
\draw[fill=color2,draw opacity=0] (axis cs:2.375,0) rectangle (axis cs:2.525,0.884444444444444);
\draw[fill=color2,draw opacity=0] (axis cs:3.375,0) rectangle (axis cs:3.525,0.890111111111111);
\addlegendimage{area legend,draw=black,fill=color3,postaction={pattern=horizontal lines}};
\draw[draw=black,fill=color3,postaction={pattern=horizontal lines}] (axis cs:0.525,0) rectangle (axis cs:0.675,0.901777777777778);
\draw[draw=black,fill=color3,postaction={pattern=horizontal lines}] (axis cs:1.525,0) rectangle (axis cs:1.675,0.940777777777778);
\draw[draw=black,fill=color3,postaction={pattern=horizontal lines}] (axis cs:2.525,0) rectangle (axis cs:2.675,0.890888888888889);
\draw[draw=black,fill=color3,postaction={pattern=horizontal lines}] (axis cs:3.525,0) rectangle (axis cs:3.675,0.901555555555555);
\end{axis}

\end{tikzpicture}}
	\end{subfigure}
	\caption{AUC values for experiment A on (a) $P_1$ and (b) $P_2$ and experiment B on (c) $P_1$ and (d) $P_2$ for the following categories of features: structural (ST), semantic (S), user-based (UB), sentiment-based (SB), and predicted (P).}\label{fig:steps}
\end{figure}
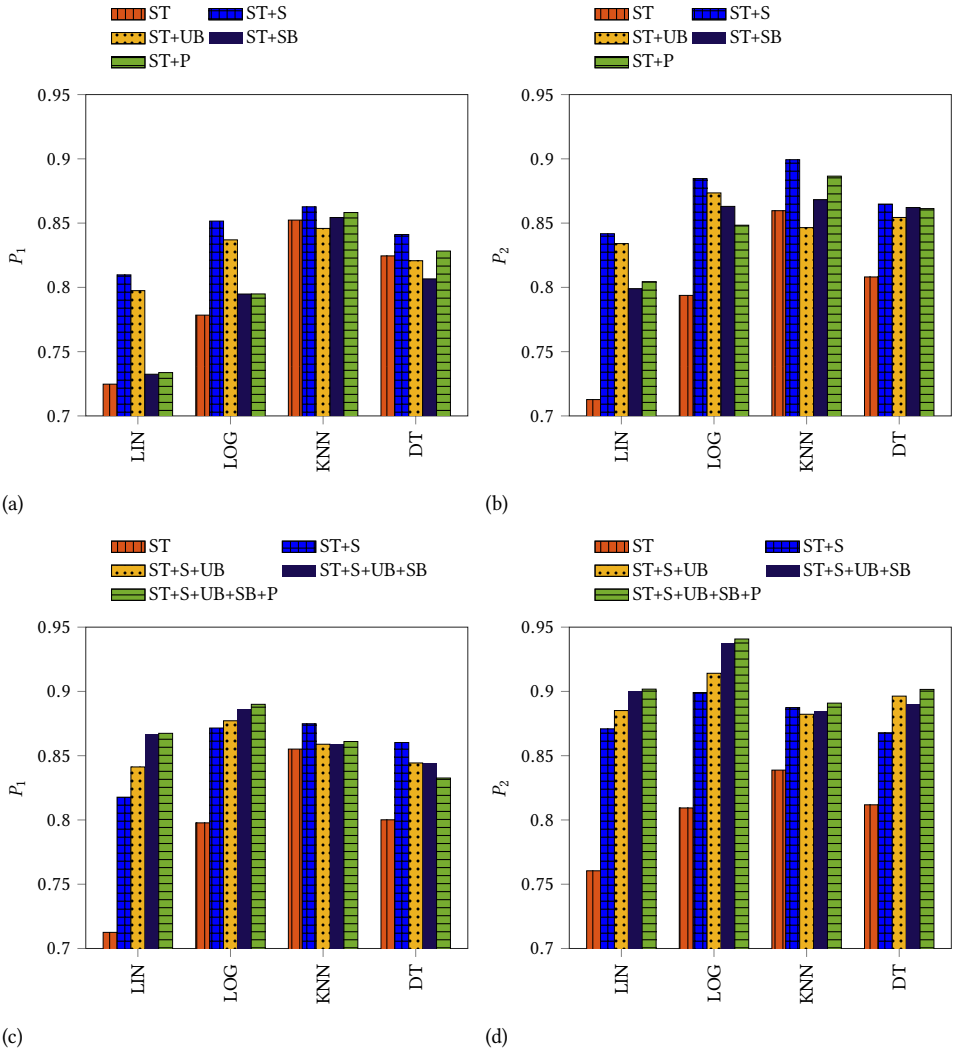

We apply our framework for the early warning to each sample $E_1$ and $E_2$, and consider their respective parameters and best classifiers, obtaining two separate sets of new features. In specifying such features, we account for the total number of predicted disputed entities and their rate, getting a total of four features, two for any of the two adopted algorithms. Under this perspective, the particularly low FP Rate of disputed entities results to be especially suited to our aim. Indeed, we may retain less information, but with a higher level of certainty. As shown in Section~\ref{sec:early_class}, to assess the potentiality of a topic to become object of fake news, our framework provides two predicted values for each entity, one for each classifier --LOG and NN. We now use these values to build two new features, that will be used in a new classifier --along with other features-- with the aim of detecting fake news. Specifically, we may define five categories of features:
	\begin{enumerate}
 	\item \textit{Structural features} i.e., related to news structure and diffusion;
	\item \textit{Semantic features} i.e., related to the textual contents of the news;
	\item \textit{User-based features} i.e., related to users' characteristics in terms of engagement and polarization;
	\item \textit{Sentiment-based features}, which refer to both the way in which news are presented and perceived by users;
	\item \textit{Predicted features}, obtained from our framework.
 	\end{enumerate}
The complete list of features used for both samples is reported in Table~\ref{tab:features_class_one}. To assess the relevance and the full predictive extent of each features category, we perform two separate five-step experiments. In the first one, experiment A, we test the performance of the introduced state-of-the-art binary classifiers by first considering structural features alone and then adding only one of the features' category at each subsequent step, for both samples $P_1$ and $P_2$. While in experiment B we consider again only structural features in the first step and then we sequentially add one of the features' category at each step. Hence for each algorithm and each experiment we get ten different trained classifiers, five for $P_1$ and five for $P_2$, where the considered categories for experiment A are: 1) structural (ST), 2) structural and semantic (ST+S), 3) structural and user-based (ST+UB), 4) structural and sentiment-based (ST+SB), 5) structural and predicted features (ST+P); while for experiment B they are: 1) structural (ST), 2) structural and semantic (ST+S), 3) structural, semantic, and user-based (ST+S+UB), 4) structural, semantic, user-based, and sentiment-based (ST+S+UB+SB), 5) structural, semantic, user-based, sentiment-based, and predicted features (ST+S+UB+SB+P).
Again, we apply different classifiers: Linear Regression (LIN), Logistic Regression (LOG), Support Vector Machine (SVM) with linear kernel, K-Nearest Neighbors (KNN) with $K=5$, Neural Network Models (NN) through the Multi-layer Perceptron L-BFGS algorithm, and Decision Trees (DT) with Gini Index. Given the asymmetry of our dataset in favor of official news sources, we re-sample the data at each step in order to get two balanced groups. We compare the results by measuring the accuracy of the predicted values through the  AUC score. In Figure~\ref{fig:steps} we report the AUC values for the two five-steps experiments. We only focus on the results of the four best performing algorithms i.e., LIN, LOG, KNN, and DT. Experiment A, shows that semantic features are the ones bringing the highest improvement w.r.t. the baseline on structural features. However, we also observe remarkable improvements for the predicted features, regardless of their small number. When looking at experiment B, we observe a significant increment in the AUC during the last step (ST+S+UB+SB+P) with respect to previous ones\footnote{The only exceptions are for k-nearest neighbors on $P_1$, where we observe the highest accuracy on step 2, and for decision trees on $P_1$, where instead we observe a decrement in the accuracy in the last step.}, and this is especially evident for logistic regression and decision trees on $P_2$. Moreover, from Figure~\ref{fig:steps}, we can see that logistic regression is the best performing algorithm and that it achieves especially high results on $P_2$. We should also notice the relative predictive power of structural features, as a matter of fact the introduction of semantic features brings, in all cases, the largest jump in the accuracy.

Table~\ref{tab:classification_results} reports classification results for the four top-ranked classifiers --LIN, LOG, KNN, and DT-- on our two samples $P_1$ and $P_2$, considering all defined features categories. We notice an overall very high level of accuracy, where the best score is $0.91$ and it is achieved by logistic regression on $P_2$, with respective precision rates in the detection of \textit{fake} and \textit{not fake} equal to $0.88$ and $0.94$. Our classifiers are generally more accurate in the detection of not fake information, however both false positive rates for fake and not fake are significantly low (especially in the LOG case), with a slightly smaller probability of falsely labeling a not fake as fake.

\vspace{0.5cm}
\begin{minipage}{0.9\linewidth}
	\captionof{table}{\textbf{Features adopted in the classification task.} 
	We count a total of $52$ features for sample $P_2$ and $44$ for sample $P_1$.}
	\label{tab:features_class_one}
	\scalebox{0.7}{
		\begin{tabularx}{1.4\linewidth}{llllll}
			\toprule
			Class & Feature name & Posts & Comments & N. of features & Sample\\
			\midrule
			&	\texttt{Number of likes/comments/shares} & \texttt{x} &- & $3$ & $P_1, P_2$ \\
			STRUCTURAL	&	\texttt{Number of likes/comments on comments} & - & \texttt{x} & $2$ &  $P_1, P_2$ \\
			&	\texttt{Average likes/comments on comments}& - & \texttt{x} & $2$ &  $P_1, P_2$ \\	\hline
			&	\texttt{Number of characters}& \texttt{x} & \texttt{x}& $2$ & $P_1, P_2$\\
			&	\texttt{Number of words}& \texttt{x} & \texttt{x}& $2$ & $P_1, P_2$  \\
			&	\texttt{Number of sentences}& \texttt{x} & \texttt{x}& $2$ &$P_1, P_2$\\
			SEMANTIC&	\texttt{Number of capital letters}& \texttt{x} & \texttt{x}& $2$ & $P_1, P_2$ \\
			&	\texttt{Number of punctuation signs}& \texttt{x} & \texttt{x}& $2$ & $P_1, P_2$\\
			&	\texttt{Average word length\footnote{\label{note_a} Average length computed w.r.t. both posts and comments.}}& \texttt{x} & \texttt{x}& $2$ &$P_1, P_2$\\
			&	\texttt{Average sentence length\footnoteref{note_a}}& \texttt{x} & \texttt{x}& $2$ & $P_1, P_2$\\
			&	\texttt{Punctuation rate\footnote{\label{note_b} Over total number of characters.}}& \texttt{x} & \texttt{x}& $2$ &$P_1, P_2$ \\	
			&	\texttt{Capital letters rate}\footnoteref{note_b}& \texttt{x} & \texttt{x} & $2$& $P_1, P_2$\\ \hline
			& \texttt{Av./Std comments to commenters}& -& -& $2$ & $P_1, P_2$\\
			& \texttt{Av./Std likes to commenters}& -& -& $2$ &$P_1, P_2$ \\
			& \texttt{Mean std likes/comments to commenters}& -& - & $2$&$P_1, P_2$ \\
			USER-BASED	& \texttt{Av./Std comments per user}& -& - & $2$& $P_1, P_2$\\
			& \texttt{Av./Std pages per user}& -& -& $2$ & $P_1, P_2$\\
			& \texttt{Total engaged users\footnote{\label{note_d} Users engaged with any of the entities detected in the post.} }& -& -& $1$ & $P_1, P_2$\\
			& \texttt{Rate of engaged users\footnoteref{note_d}} & -& -& $1$& $P_1, P_2$\\			\hline
			& \texttt{Sentiment score}& \texttt{x} & -  & $1$& $P_1, P_2$\\ 
			& \texttt{Av./Std comments' sentiment score}& - & \texttt{x} & $2$ & $P_2$\\ 
			&	\texttt{Rate positive/negative comments}& - & - & $2$ &  $P_2$ \\
			SENTIMENT-BASED	& \texttt{Number of positive over negative comments}& \texttt{x} & - & $1$ & $P_2$  \\
			& \texttt{Mean/Std presentation distance
			}& \texttt{x} & - & $2$ &$P_1, P_2$ \\ 
			& \texttt{Number/Rate of captivating entities\footnote{\ Entities for which $\kappa(e) = 1$ (see Section~\ref{sec:feat} for details).}}& \texttt{x} & - & $2$ &$P_2$ \\ 
			& \texttt{Av. response distance
			}& - & - & $1$ &$P_2$ \\			\hline
			& \texttt{Numb. of pred. D entities\footnote{\label{note_h} Our framework for the early warning allows to classify an entity as disputed or not. Here we consider disputed entities predicted through logistic regression (LOG) and neural networks (NN), since they proved to be the best performing algorithms (see Section~\ref{sec:early_class} for details).} (LOG, NN)}& - & -  & $2$& $P_1, P_2$\\
			PREDICTED	& \texttt{Rate of pred. D entities\footnoteref{note_h} (LOG, NN)}& - & - & $2$ & $P_1, P_2$\\
			\bottomrule
			\end{tabularx}}
\end{minipage}
\vspace{3mm}

We select the best performing features on the basis of their AUC score, employing the forward stepwise features selection (see Section~\ref{sec:early_class}). In Table~\ref{tab:best_features_one}  we report the 16 best performing features for both samples $P_1$ and $P_2$. We note that predicted features are among the best performing features in both cases, however structural and semantic features are the most represented. Also, the mean presentation distance appears among the best in both columns. We may deduce that the newly introduced sentiment-base and predicted features are extremely relevant for the purpose of fake news identification. Moreover, the potential influential character of the commenters, embodied in the average number of comments to the commenters, is either the first or the second best performing features, underling the, often neglected, primary role of intermediate nodes in the diffusion of fake news.

\begin{table}[tb]
	\centering
	\caption{\textbf{Classification results.} We report the performances for the 4 best performing algorithm (LIN, LOG, KNN, DT). The first reported values refer to $P_2$, as in that case we have more features for the classification and we also get better results, while those for $P_1$ are in parentheses. W. Avg. denotes the weighted average across the two classes.}
	\label{tab:classification_results} 
	\scalebox{0.9}{	\begin{tabular}{lllllllll}
			\toprule
			& AUC &Accuracy&M. Abs. Err.&	  &Precision& Recall& FP Rate&F1-score\\
			\midrule
		     & & & &	{\textit{Not Fake}}  & $0.91 (0.85)$ & $0.90 (0.84)$  &$0.11 (0.15) $& $0.91 (0.85)$\\ 
	        {\textit{LIN}} & 0.90 (0.87) & $0.90 (0.84)$& $0.10 (0.16)$ &	{\textit{Fake}}  &$0.88 (0.84)$ & $0.90 (0.85)$ & $0.10 (0.16)$& $0.89 (0.84)$\\
	        & & & &	{\textit{W. Avg.}}   & $0.90 (0.84)$ & $0.90 (0.84)$&$0.11 (0.16)$ & $0.90 (0.84)$\\	        
	        \midrule
	        & & & &	{\textit{Not Fake}}  & $0.94 (0.90)$ & $0.90 (0.87) $& $0.07 (0.10) $ & $0.92 (0.88)$\\
	        {\textit{LOG}} & \textbf{0.94 (0.89)} & $\textbf{0.91 (0.88)}$& $\textbf{0.09 (0.12)}$&	{\textit{Fake}}   &$0.88 (0.87)$ & $0.93 (0.90)$ & $0.10 (0.13)$& $0.90 (0.88)$\\
	        & & & &	{\textit{W. Avg.}}  & $0.91 (0.88)$ & $0.91 (0.88)$ & $0.08 (0.12)$& $0.91 (0.88)$\\
	        \midrule
	        & & & &	{\textit{Not Fake}}  & $0.90 (0.82)$ & $0.86 (0.82)$ &$0.11 (0.18) $ & $0.88 (0.82)$\\ 
	        {\textit{KNN}} & 0.89 (0.86) & $0.87 (0.82)$& $0.13 (0.18)$ &	{\textit{Fake}} &$0.84 (0.81)$ & $0.89 (0.82)$ & $0.14 (0.18)$ & $0.87 (0.82)$\\
	        & & & &	{\textit{W. Avg.}}& $0.87 (0.82)$ & $0.87 (0.82)$  &$0.13 (0.18)$ & $0.87 (0.82)$\\
	        \midrule   
	        & & & &	{\textit{Not Fake}}   &$0.92 (0.86)$ & $0.86 (0.83)$  &$0.09 (0.14) $ & $0.89 (0.84)$\\ 
	        {\textit{DT}}  & 0.90 (0.83) & $0.89 (0.85)$& $0.11 (0.15)$&	{\textit{Fake}}    & $0.85 (0.83)$ & $0.91 (0.86)$ &$0.14 (0.17)$& $0.88 (0.85)$\\
	        & & & &	{\textit{W. Avg.}}   & $0.89 (0.85)$ & $0.89 (0.84)$ &$0.12 (0.16)$  & $0.89 (0.84)$\\
	        \bottomrule
	\end{tabular}}
\end{table}	

\begin{table}[tb]
	\centering
	\caption{\textbf{Best performing features for post classification.}  Labels $P$ and $C$ indicate if the feature is computed w.r.t. either posts or comments. For each sample we report the feature and its respective category.}
	\label{tab:best_features_one}
	\scalebox{0.85}{	\begin{tabular}{lll|ll}
			\toprule
			&	$P_1$& &$P_2$&\\
			\midrule
			& \textit{Feature}&\textit{Cat.}& \textit{Feature}&\textit{Cat.}\\
			\midrule
			$1$ & \textbf{Av. numb. of comments to comm.ers} & UB & Number of words (P)& S \\ 
			$2$ & \textbf{Numb. of predicted entities disputed (NN)}& P& \textbf{Av. numb. of comments to comm.ers} & UB\\ 
			$3$ &   Number of words (P)& S&  Number of likes (P)& ST\\ 
			$4$ &   Number of likes (P)& ST &  Number of shares (P)&ST\\ 
			$5$ &   Number of shares (P)& ST& Capital letters rate (P)&S\\ 
			$6$ & \textbf{Std numb. of likes to comm.ers} & UB &Number of comments (C)&ST\\ 
			$7$ & Number of capital letters (P)& S &Number of comments (P) &ST\\ 
			$8$ & Number of punctuation signs (P)& S&Av. comments per user&UB\\ 
			$9$ &  Std numb. of comments per user&UB &Number of sentences&S \\ 
			$10$ &  Std sentiment score (C)& SB& Number of punctuation signs&S\\ 
			$11$ &   Number of characters (CP)& S &\textbf{Numb. of predicted disputed entities (NN)} &P\\ 
			$12$ &  Av. sentence length (P) & S&\textbf{Mean presentation distance} &SB \\ 
			$13$ & \textbf{Mean presentation distance} &SB & Av. numb. of comments to comm. & ST\\ 
			$14$ &  Number of comments (C)& ST &Rate of polarized users&UB \\ 
			$15$ &  \textbf{Rate of predicted disputed entities (LOG)}& P &\textbf{Std presentation distance} &SB \\ 
			$16$ &  \textbf{Rate of predicted disputed entities (NN)}& P &Av. pages per user & UB  \\ 
			\bottomrule
	\end{tabular}}
\end{table}	

\newpage
\section{Conclusions}\label{conclusion}
In this article, we presented a general framework for a timely identification of polarizing content that enables to 1) ``predict'' future fake news topics on social media, and 2) build a classifier for fake news detection. We validated the performances of our methodology on a massive dataset of official news and hoaxes on Facebook, however an extension to other social media platforms is straightforward. Our analysis shows that a deep understanding of users' behavior and polarization is crucial when dealing with the problem of misinformation. To our knowledge, this is the first attempt towards the early detection of possible future topics for fake news, still not without limitations --mainly due to the fact that fake or unsubstantiated information is often diffused even by official newspapers. When dealing with a complex issue such as massive digital misinformation, special caution is required. However, our results are promising and bode well for a system enabled for monitoring information flow in real time and issuing a warning about delicate topics. In this direction, our approach could represent a pivotal step towards the smoothing of polarization on online social media.

\bibliographystyle{IEEEtran}
\bibliography{acmart}
\end{document}